\documentclass{bmcart}

\RequirePackage{graphicx}
\RequirePackage{hyperref}
\usepackage[utf8]{inputenc} 

\usepackage[acronym]{glossaries}
\usepackage{paralist}
\usepackage{listings,,multicol}

\makeglossaries

\startlocaldefs
\endlocaldefs

\usepackage{xcolor}

\newif\ifanonymize

\anonymizefalse

\newcommand{\anon}[1]{\ifanonymize{Removed for double-blind review process}\else{#1}\fi}
\newcommand{\anc}[1]{\ifanonymize{}\else{#1}\fi}

\usepackage{hyphenat}

\begin{document}

\newacronym[\glslongpluralkey={Smart Grids}]{sg}{SG}{Smart Grid}
\newacronym[\glslongpluralkey={Distributed Energy Resources}]{der}{DER}{Distributed Energy Resource}
\newacronym[\glslongpluralkey={Information and Communication Technologies}]{ict}{ICT}{Information and Communication Technology}
\newacronym{fdi}{FDI}{False Data Injection}
\newacronym{scada}{SCADA}{Supervisory Control and Data Acquisition}
\newacronym[\glslongpluralkey={Master Terminal Units}]{mtu}{MTU}{Master Terminal Unit}
\newacronym[\glslongpluralkey={Cyber-Physical Systems}]{cps}{CPS}{Cyber-Physical System}
\newacronym[\glslongpluralkey={Industrial Control Systems}]{ics}{ICS}{Industrial Control System}
\newacronym{hmi}{HMI}{Human Machine Interface}
\newacronym[\glslongpluralkey={Programmable Logic Controllers}]{plc}{PLC}{Programmable Logic Controller}
\newacronym[\glslongpluralkey={Intelligent Electronic Devices}]{ied}{IED}{Intelligent Electronic Device}
\newacronym[\glslongpluralkey={Remote Terminal Units}]{rtu}{RTU}{Remote Terminal Unit}
\newacronym{iec104}{IEC-104}{IEC 60870-5-104}
\newacronym{apdu}{APDU}{Application Protocol Data Unit}
\newacronym{apci}{APCI}{Application Protocol Control Information}
\newacronym{asdu}{ASDU}{Application Service Data Unit}
\newacronym[\glslongpluralkey={Information Objects}]{io}{IO}{Information Object}
\newacronym{cot}{COT}{Cause Of Transmission}
\newacronym{ot}{OT}{Operational Technology}
\newacronym{mitm}{MITM}{Man-in-the-Middle}
\newacronym{fdi}{FDI}{False Data Injection}
\newacronym{dmz}{DMZ}{Demilitarized Zone}
\newacronym[\glslongpluralkey={Intrusion Detection Systems}]{ids}{IDS}{Intrusion Detection System}
\newacronym{siem}{SIEM}{Security Information and Event Management}
\newacronym{mv}{MV}{Medium Voltage}
\newacronym{lv}{LV}{Low Voltage}
\newacronym{cdss}{CDSS}{Controllable Distribution Secondary Substation}
\newacronym[\glslongpluralkey={Battery Storage Systems}]{bss}{BSS}{Battery Storage System}
\newacronym{pv}{PV}{Photovoltaic}
\newacronym[\glslongpluralkey={Measuring Points}]{mp}{MP}{Measuring Point}
\newacronym{dsc}{DSC}{Dummy SCADA Client}
\newacronym{fcli}{FCLI}{Fronius CL inverter}
\newacronym{fipi}{FIPI}{Fronius IG+ inverter}
\newacronym{sii}{SII}{Sunny Island inverter}
\newacronym{tls}{TLS}{Transport Layer Security}
\newacronym{actcon}{ActCon}{Activation Confirmation}
\newacronym{actterm}{ActTerm}{Activation Termination}
\newacronym{rtt}{RTT}{Round Trip Time}
\newacronym{c2}{C2}{Command and Control}
\newacronym{dst}{DST}{Dempster Shafer Theory}
\newacronym{ec}{EC}{Event Correlator}
\newacronym{sc}{SC}{Strategy Correlator}
\newacronym{ioa}{IOA}{Information Object Address}
\newacronym[\glslongpluralkey={Indicators of Compromise}]{ioc}{IoC}{Indicator of Compromise}
\newacronym[\glslongpluralkey={False Positives}]{fp}{FP}{False Positive}
\newacronym[\glslongpluralkey={False Negatives}]{fn}{FN}{False Negative}
\newacronym[\glslongpluralkey={True Positives}]{tp}{TP}{True Positive}
\newacronym[\glslongpluralkey={True Negative}]{tn}{TN}{True Negative}
\newacronym{dpi}{DPI}{Deep Packet Inspection}
\newacronym{fpr}{FPR}{False Positive Rate}
\newacronym{wan}{WAN}{Wide Area Network}
\newacronym{scl}{SCL}{Substation Configuration Language}
\newacronym{gim}{GIM}{Graph-based Infrastructure Model}
\newacronym[\glslongpluralkey={Specification-based IDSs}]{sids}{SIDS}{Specification-based IDS}
\newacronym[\glslongpluralkey={Automata Models}]{am}{AM}{Automata Model}
\newacronym{sb}{SB}{Specification Base}

\begin{frontmatter}

\begin{fmbox}
\dochead{Research}


\title{On Specification-based Cyber-Attack Detection in Smart Grids}


\author[addressref={aff1,aff2},email={\anon{o.sen@iaew.rwth-aachen.de}}]{\anon{\inits{ÖS}\fnm{Ömer} \snm{Sen}}}
\author[addressref={aff1,aff2},email={\anon{dennis.van.der.velde@fit.fraunhofer.de}}]{\anon{\inits{DV}\fnm{Dennis} \snm{van der Velde}}}
\author[addressref={aff1},email={\anon{maik.luehman@rwth-aachen.de}}]{\anon{\inits{ML}\fnm{Maik} \snm{Lühman}}}
\author[addressref={aff1},email={\anon{florian.spruenken@rwth-aachen.de}}]{\anon{\inits{FS}\fnm{Florian} \snm{Sprünken}}}
\author[addressref={aff1,aff2},email={\anon{i.hacker@iaew.rwth-aachen.de}}]{\anon{\inits{IH}\fnm{Immanuel} \snm{Hacker}}}
\author[addressref={aff1,aff2},email={\anon{a.ulbig@iaew.rwth-aachen.de}}]{\anon{\inits{AU}\fnm{Andreas} \snm{Ulbig}}}
\author[addressref={aff1,aff2},email={\anon{michael.andres@fit.fraunhofer.de}}]{\anon{\inits{MA}\fnm{Michael} \snm{Andres}}}
\author[addressref={aff3,aff4},email={\anon{henze@cs.rwth-aachen.de}}]{\anon{\inits{MH}\fnm{Martin} \snm{Henze}}}



\address[id=aff1]{\anon{\orgname{High Voltage Equipment and Grids, Digitalization and Energy Economics, RWTH Aachen University}, \street{Schinkelstraße 6}, \postcode{52062} \city{Aachen}, \cny{Germany}}}
\address[id=aff2]{\anon{\orgname{Fraunhofer Institute for Applied Information Technology FIT}, \street{Schloss Birlinghoven, Konrad-Adenauer-Straße}, \postcode{53757} \city{Sankt Augustin}, \cny{Germany}}}
\address[id=aff3]{\anon{\orgname{Security and Privacy in Industrial Cooperation, RWTH Aachen University}, \street{Ahornstraße 55}, \postcode{52074} \city{Aachen}, \cny{Germany}}}
\address[id=aff4]{\anon{\orgname{Fraunhofer Institute for Communication, Information Processing and Ergonomics FKIE}, \street{Fraunhoferstraße 20}, \postcode{53343} \city{Wachtberg}, \cny{Germany}}}


\end{fmbox}


\begin{abstractbox}

\begin{abstract} 
The transformation of power grids into intelligent cyber-physical systems brings numerous benefits, but also significantly increases the surface for cyber-attacks, demanding appropriate countermeasures. 
However, the development, validation, and testing of data-driven countermeasures against cyber-attacks, such as machine learning-based detection approaches, lack important data from real-world cyber incidents.
Unlike attack data from real-world cyber incidents, infrastructure knowledge and standards are accessible through expert and domain knowledge.
Our proposed approach uses domain knowledge to define the behavior of a smart grid under non-attack conditions and detect attack patterns and anomalies.
Using a graph-based specification formalism, we combine cross-domain knowledge that enables the generation of whitelisting rules not only for statically defined protocol fields but also for communication flows and technical operation boundaries.
Finally, we evaluate our specification-based intrusion detection system against various attack scenarios 
and assess detection quality and performance.
In particular, we investigate a data manipulation attack in a future-orientated use case of an IEC 60870-based SCADA system that controls distributed energy resources in the distribution grid.
Our approach can detect severe data manipulation attacks with high accuracy in a timely and reliable manner.
\end{abstract}


\begin{keyword}
\kwd{Cyber Security}
\kwd{Cyber Physical Systems}
\kwd{Intrusion Detection Systems}
\end{keyword}

\end{abstractbox}

\end{frontmatter}

\section*{Introduction} \label{sec:introduction}
The paradigm shift that is taking place in the energy sector as part of the energy transition due to the increasing penetration of \glspl{der} poses new challenges for grid operators, especially at the distribution grid level~\cite{1_ourahou2020review}.
To meet these challenges, a more active role of the distribution grid operator is required through increased expansion of sensors and actuators, which provide telecontrol connections via \glspl{ict} to resources such as controllable \glspl{der}~\cite{2_smartrgird2020book}.
This transformation from traditional grid structures to intelligent networked energy information systems -- \glspl{sg} -- using \glspl{ict} not only opens up new opportunities and solutions to master the energy transition, but also new dangers that threaten resilience and cyber-security~\cite{3_van2020methods}.

Reliable and secure grid operation increasingly depends on properly functioning communication technologies and processes due to the high penetration of \glspl{ict}, making it more vulnerable to failures and cyber-attacks~\cite{4_klaer_sgam_2020}.
In particular in the context of \gls{ics}, which also includes process networks of power grids, a threat landscape against cyber-attacks becomes apparent, which is essentially characterized by a long lifetime of assets and the use of legacy components with limited security mechanisms~\cite{5_eder2017cyber}:
In 2015, unauthorized third parties exploited these vulnerabilities to gain control of remotely controlled equipment, such as circuit breakers, to disrupt the power supply of more than 225,000 customers~\cite{6_case2016analysis}.

To counter new threats such as cyber-attacks, and to protect basic security objectives, i.e., confidentiality, availability, and integrity, cyber-security countermeasures are required, which are divided into preventive and reactive or active and passive measures.
Various guidelines and standards, e.g., the IEC 62531 series of standards~\cite{30_wg152016iec}, specify countermeasures such as the use of cryptography and authentication procedures in the telecontrol protocols.
However, given the long-standing legacy devices with performance and resource constraints, countermeasures with high performance overhead involve high expenditures and costs to implement upgrades or workarounds~\cite{8_tanveer2020secure}.
More passive and reactive security measures are network-based \glspl{ids}, which passively record communication traffic and perform attack detection within the process networks at selected points~\cite{wolsing2022ipal}.

Intrusion detection methods can be broadly divided into blacklisting, in which observations are compared with known attack signatures, and whitelisting, in which observations are compared with the established understanding of the system's characteristics under normal conditions~\cite{11_krause2021cybersecurity}.
For process networks with deterministic network structures and physically plausible verifiable payloads, whitelisting is a promising methodology to detect attacks or anomalies without prior knowledge of patterns and signatures~\cite{9_eckhart2018specification}.
Furthermore, the impact of missing or hard-to-access data on attacks against power grids on the effectiveness of detection methods can be reduced, as whitelisting approaches do not primarily require such data.

A challenge in applying a whitelisting approach is the need for a holistic capture of the characteristics of the system and its formalism, which includes technical and operational specifications of the infrastructure, as well as the behavior of the devices under normally defined states.
Possibilities for this capture can be machine-learning-based or domain-specific knowledge-based approaches~\cite{11_krause2021cybersecurity}.
The first approach is essentially characterized by the automated generation of a model in defined learning periods, which is trained and generated from recorded communication data~\cite{10_baraneetharan2020role}.
In particular, in combination with \gls{dpi}, industrial protocols such as \gls{iec104} are decomposed into relevant fields to collect training data for models to detect the derivation of a standard pattern and suspicious processes in the form of anomalies~\cite{mochalski2020cybersicherheit}.
However, this may imply a potential vulnerability to data manipulation during the learning phases and incompleteness of non-observable legitimate situations such as maintenance in the training datasets.
In the second approach, specifications are defined that are used as a set of rules to define the characteristics of the system under normal conditions to detect anomalies.
E.g., the IEC 61850 standard, which is mainly used in substations, describes data models and communication parameters in a format known as \gls{scl}, which can provide the \gls{sb} for normal conditions~\cite{hokama2020cybersecurity}.
However, the thoroughly available data prescribed by the standard's data model is not applicable to all industrial protocols, such as \gls{iec104}.
Thus, the \gls{sids} approach requires proper domain knowledge deposited and validated, but potentially achieves higher precision rates compared to machine-learning-based \gls{ids} approaches~\cite{12_verma2020machine,kus2022false}.

However, the holistic capture of infrastructure specification and behavior of components in normal states requires high manual efforts in bundles of cross-domain knowledge and their maintenance, resulting in technically complex implementations without suitable accessible formalization.
We identify the following challenges:
\begin{itemize}
	\item Concentration of dispersed domain-specific knowledge in a holistic \gls{sb} description of the infrastructure.
	\item Automated extraction of detection rules from infrastructure knowledge to detect anomalies and suspicious events.
	\item Provision of explanations for issued alerts through coherent rule matching of infrastructure knowledge and alert generation.
	\item Maintain high flexibility in detection capabilities through modular enrichment of infrastructure knowledge.
\end{itemize}

To address these challenges, we propose an approach for a \gls{sids}, which, supported by an infrastructure specification and \glspl{am} for component behavior w.r.t.\ communication flows, captures characteristics of the \gls{sg} for cyber-attack detection.
More precisely, our contributions are:
\begin{enumerate}
	\item We identify relevant domain knowledge for the \gls{sids} of cyber-attacks and intrusions in \glspl{sg} by extracting domain-specific data based on a \gls{gim} approach (\nameref{subsec:framework_specification}).
 	\item We present and describe a structured approach for detecting anomalies in communication behavior in \glspl{sg} process networks that uses a holistic \gls{gim} as a \gls{sb} and \glspl{am} for flow consistency checks (\nameref{subsec:framework_deeppacketinspect}).
	\item We demonstrate and discuss the performance of our proposed approach against different attack scenarios in a physical testbed by evaluating the detection quality and performance within the scenarios (\nameref{sec:result}).
\end{enumerate}

\section*{Smart Grid \& Process-awareness in Detection} \label{sec:background}
As the basis of our work, we describe the infrastructure specification of process networks based on \gls{scada}, discuss their security and related research work.

\subsection*{SCADA-based Process Networks} \label{subsec:background_scada}
\begin{figure}
\centering
\includegraphics[width=0.95\linewidth]{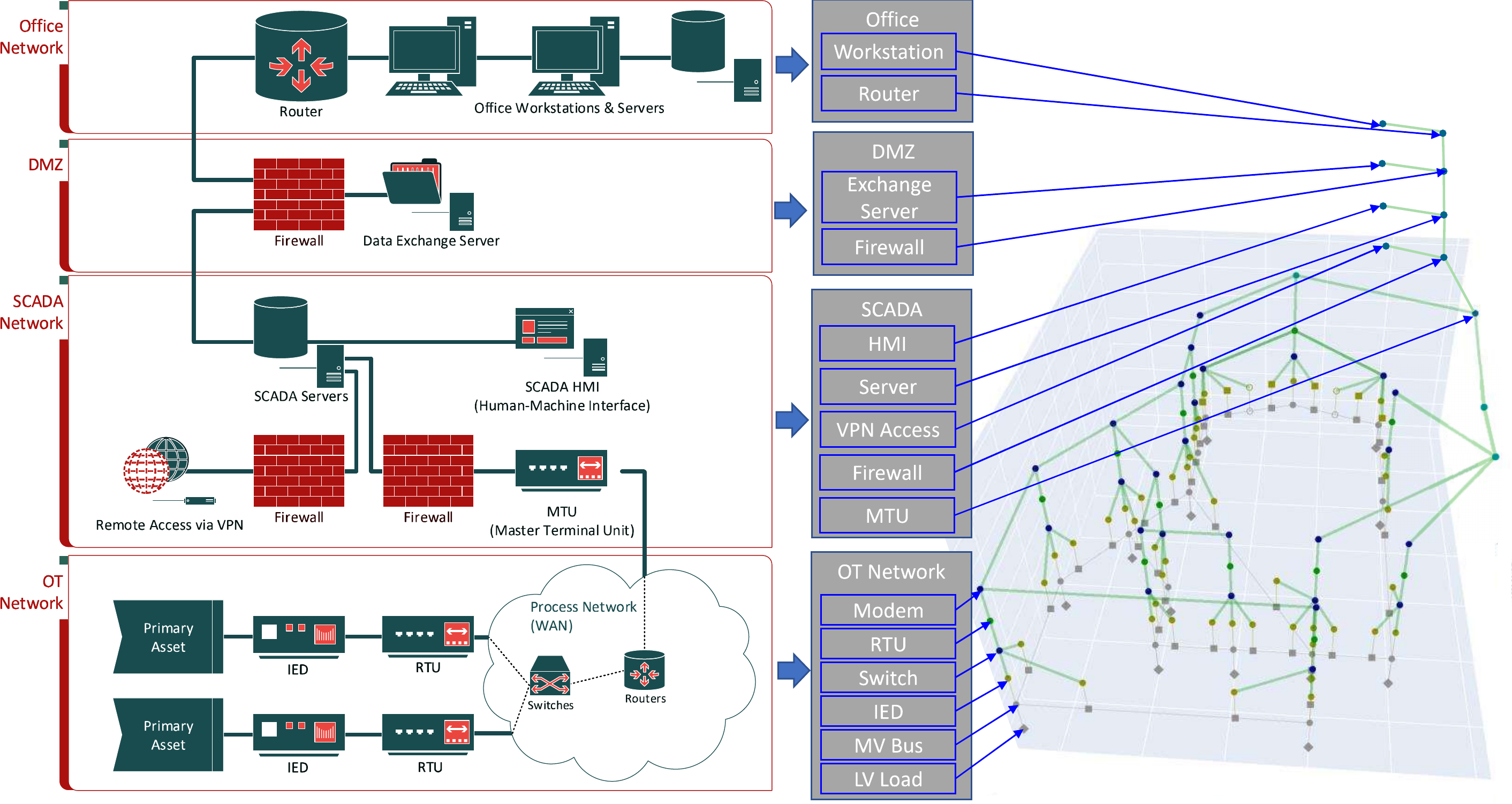}
\caption{Illustration of the \gls{sg} infrastructure based on the Purdue reference model, representing a future-oriented \gls{scada} system that is connected to the primary equipment via dedicated and/or public communication infrastructure\anc{~\cite{26_vandervelde2021towards}}. The right side represents the graph-based formalism of the infrastructure as a \gls{sb}\anc{~\cite{4_klaer_sgam_2020}}.}
\label{fig:smartgrid_spec}
\end{figure}
Based on the Purdue reference model~\cite{13_williams1994purdue}, process networks based on \gls{scada} in \glspl{sg} can be described as hierarchical control structures consisting of primary technology, secondary technology, and \gls{ict}~\cite{2_smartrgird2020book} as presented in Figure~\ref{fig:smartgrid_spec}.
Divided into several levels, the primary technology, such as circuit breaker switches, is monitored and controlled utilizing secondary technology, such as sensors and actuators.

At the lowest level of the control hierarchy, the infrastructure is divided into the \gls{ot} network (equipment and systems with telecontrol connection through \gls{wan}).
Directly connected to the \gls{ot} network is the \gls{scada} network (control systems and communication stations with operator stations).
Connected via firewalls, a security-controlled level, the \gls{dmz}, is present (logically segmented area between corporate office and \gls{scada} network).
Finally, the corporate network is connected to \gls{dmz}.

The primary facilities are connected to the \gls{ot} network through \glspl{ied} used for control and measurement tasks, aggregated within the control hierarchy by \glspl{rtu}.
Data are then forwarded to the \gls{scada} system through the \gls{ot} network using appropriate \gls{ot} protocols such as \gls{iec104}~\cite{22_IEC104_2006} or Modbus~\cite{23_Modbus_2012}.
Within the \gls{ot} network, the \gls{mtu} counterpart to the \gls{rtu} acts as a gateway for the \gls{scada} system.

\subsection*{Cyber-Security in Process Networks} \label{subsec:background_cybersec}
In the European energy sector, the \gls{iec104} protocol is often used to monitor and control geographically widely distributed processes~\cite{21_matouvsek2017description}.
\gls{iec104} as a legacy industrial protocol does not provide security features such as encryption or authentication~\cite{30_wg152016iec}.
Therefore, without an encryption or authentication mechanism, unauthorized third parties can intercept critical \gls{iec104} traffic and potentially endanger the grid.
E.g., the attacker can intercept existing communication channels by a \gls{mitm} attack or establish new connections to manipulate the traffic.
Thus, the attacker would be able to read, modify, inject, or discard new or sent messages between the intercepted or newly connected endpoints~\cite{31_yang2012man}.

To address the critical security issues within the process network, especially the legacy protocols, the IEC 62351 standard discusses new security principles and requirements.
E.g., the IEC 62351 standard requires secure end-to-end communication using the \gls{tls} protocol, which provides secure key exchange, encryption, and authentication~\cite{32_iec62351_2018}.
However, large-scale implementation and adaptation of the new standards in traditional process networks are hampered by the large number of resource-constrained devices, which may jeopardize service availability.
These approaches can overwhelm resource-limited field devices such as \glspl{rtu} or \glspl{ied} and cause higher communication latency, preventing \gls{scada} applications from meeting real-time requirements~\cite{8_tanveer2020secure}.

Different studies investigated the performance issues caused by \gls{tls} protocol integration, and negative impacts on the performance of industrial protocol communications (e.g., IEC 61850, \gls{iec104}) have been observed~\cite{34_todeschini2020securing}.
Power grids often contain performance-limited assets with long depreciation periods that cannot be replaced or upgraded without high costs, which require legacy compliant solutions~\cite{35_castellanos2017legacy}.
\gls{ids} can provide passive security via detective capabilities to identify possible attack indicators or anomalies that do not actively interfere with the process network~\cite{36_fernandes2019comprehensive}.
There are several \gls{ids} approaches to identify potentially suspicious events, either by comparing observation with knowledge that represents normal system behavior, or by directly comparing the signature with known classified attacks~\cite{37_zuech2015intrusion}.

However, the latter approach requires attack data for detection, which limits flexibility in detecting unknown attacks such as zero-days~\cite{akshaya2019study}.
Moreover, comparing observations with known normal conditions based on trained models using data-driven machine learning approaches also has the disadvantage of low accuracy and limitation due to the scenarios included in the training data~\cite{38_khraisat2019survey}.
Therefore, a specification-based approach that relies on verified expert knowledge has the potential to provide high accuracy in detection and reduce the flexibility constraint by relying on domain-specific knowledge.
The challenge with \gls{sids} approaches is to provide a standardized \gls{sb} for different SG use cases, based on which anomaly detection conditions can be automatically derived.
Therefore, in this paper, we present a \gls{sids} that uses a defined \gls{gim} to automatically derive the set of rules.

\subsection*{Related Work} \label{subsec:background_relatedwork}
Many studies and research works have investigated detection mechanisms based on process-awareness of \gls{cps} for their suitability as \gls{ids}.

One of these research directions involves addressing process-aware \glspl{ids} that evaluates the attractiveness and criticality of \gls{ics} devices that underlie industrial processes that could be modified to achieve adversary goals~\cite{15_cook2017industrial}.
On this basis, the necessary signatures or heuristics that an adversary will leave as traces in its compromise attempt are identified.
Another research approach uses the degradation and functionality features of control signals to extract the meaning of the process of commands and determine the nondegradation pattern of the control signal within the action chain~\cite{16_escudero2018process}.
The goal is to detect the unlegitimacy of the control signal issued by \gls{ied} to the action chains before it controls the equipment.

Toward a holistic coverage of \glspl{cps}, there are approaches that replicate the program states from physical devices to their digital twins using passive data sources and system specifications~\cite{9_eckhart2018specification}.
Using stimuli and replication in a virtual environment, detailed testing is enabled in the context of \gls{ids}.
More advanced research approaches address cyber-attack classification prepared in laboratory experiments and performed in tests to design various \gls{ids} rules~\cite{18_mohan2020distributed}.
The approach is based on rule generation algorithms in a distributed architecture to accommodate \gls{scada} traffic.

Another approach pursues process-aware \gls{ids} by modeling \gls{ics}/\gls{scada} communication using probabilistic automata~\cite{19_matouvsek2021efficient}.
The model represents normal communication with a small number of states and edges whose semantics are extracted from the headers of the protocol and detect state-based anomalies.
In the context of state interpolations in the automaton, an approach is presented that uses a combination of fuzzy interpolation with fuzzy automata~\cite{20_almseidin2019fuzzy}.
Using automata theory and the fuzzy system for reasoning as part of the detection mechanism, a state transition rule base method is implemented to detect attacks.

Regarding the anomaly detection methods for \gls{iec104}, some multivariate access control and outlier detection approaches have been proposed using extracted packet information and communication statistics through Scapy~\cite{rohith2018scapy} and CICFlowMeter~\cite{lashkari2017cicflowmeter} for anomaly detection~\cite{grammatikis2020anomaly}.
In the area of statistically based anomaly detection on \gls{iec104}, the work in~\cite{burgetova2021anomaly} presents a 3-value detection method that independently compares the number of packets transmitted in three consecutive time windows against a statistical profile and reports anomalies when a deviation from the specified range is detected.
To address the problem of missing labeled data, the work of~\cite{anwar2021comparison} explores the use of unsupervised machine learning on \gls{iec104}, in particular, one-class support vector machines, isolation forest, histogram-based outlier detection, and k-nearest neighbor are investigated.

When addressing security issues within the protocol \gls{iec104}, research in~\cite{scheben2017status} examined the detection qualities of a machine learning-based detection system compared to a misuse-based system such as Snort~\cite{caswell2004snort}.
The result undermines the flexibility-accuracy dilemma described earlier, where the misuse-based system has high accuracy but low flexibility, whereas the machine learning-based system has higher flexibility but lower accuracy.
Furthermore, the challenge with automated machine learning-based detection systems also lies in their explainability~\cite{dang2021improving}, which challenges the plausibility check of the output~\cite{holzinger2020measuring}.

Although the proposed approaches provide different mechanisms to combine process knowledge with cyber-security, they still require significant additional analytical resources to provide the necessary information for their functionality.
E.g., in addition to the infrastructure specification for which the operator can provide necessary knowledge, additional efforts must be made to develop an understanding of likely attack targets, details about stimuli, statistical data, or vendor-specific technical specifications such as equipment degradation that are often inaccessible.
Therefore, our approach is entirely based on the utilization of domain-specific knowledge, which is accessible from standards and infrastructure knowledge from grid operators.
While the implementation of \glspl{am} enables the detection of inconsistencies within processes and flows, it does not take into account the semantics of the data points involved in the traffic.
Therefore, the intrusion detection capabilities of our proposed \gls{sids} rely not only on automata-based detection, but also on semantic verification of the data points.
The \gls{gim} encapsulates the semantics of the data points, which is part of advanced detection.
Through the holistic formulation of a graph-based specification foundation that provides the required overall understanding of the process, semantics, and communication of \gls{sg}, we design a process-aware \gls{sids}.

\section*{Specification-based Intrusion Detection} \label{sec:framework} 
In this section, our approach of a \gls{sids} for \glspl{sg} is presented.

In general, our approach is based on \gls{sb} derived from \gls{gim} that encapsulates domain-specific knowledge.
Using \gls{sb}, monitored network traffic is checked against \gls{sb}, with a violation resulting in a specific and explainable alert.
Furthermore, the consistency of communication behavior is checked against protocol-specific \glspl{am} to ensure that the industrial packet flows comply with the state transitions.
Thus, intrusion detection is performed using a mixture of approaches leveraging the domain knowledge of a \gls{gim}.

\subsection*{Framework Overview} \label{subsec:framework_overview}
\begin{figure}
    \centerline{\includegraphics[width=\columnwidth]{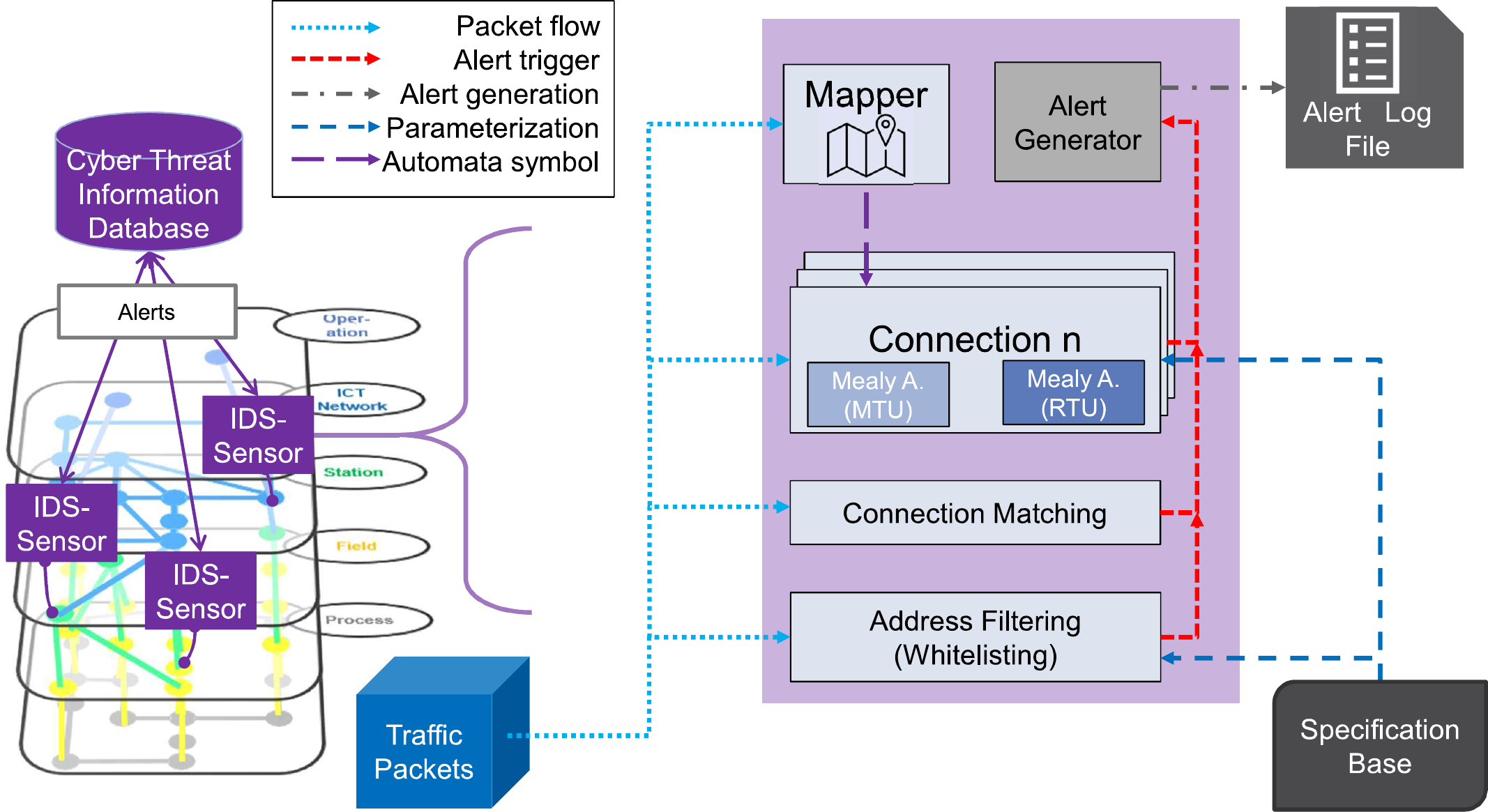}}
    \caption{This illustration presents the \gls{sids} approach to observe network traffic based on specifications and indication of traffic, which is a hybrid approach of specification-based rule matching and behavior consistency checking via a Mealy automata.}
    \label{fig:framework_overview}
\end{figure}
Our proposed \gls{sids} (cf. Figure~\ref{fig:framework_overview}) is a network-based \gls{sids} that checks for the presence of malicious content in the various layers of the network protocol.
The \gls{sids} detects anomalies based on the captured traffic by checking information from the packet headers and payloads within the industrial protocol stack (e.g., Ethernet/IP/TCP/\gls{iec104}) and discrepancies in the packet flow using Mealy automata~\cite{grigorchuk2000automata}.
The Mealy automaton has no accepting states, and the sequence of outputs from the sequence of transitions leads to a reactive system with transitions as its core and is therefore a more appropriate model for protocols~\cite{bieniasz2016towards}.
Based on a variation of Angluin's L* algorithm that generates Mealy automata~\cite{27_bollig2010libalf}, the state machine model is used to perceive the communication behavior.

In this paper, we focus on the \gls{iec104} protocol, which is a widely used standard in the European energy sector for monitoring and controlling tasks in TCP/IP-based networks~\cite{21_matouvsek2017description}.
However, the \gls{sids} is not limited to the contents of the application layer of \gls{iec104} traffic, but considers all layers that are included in typical TCP/IP-based \gls{iec104} packets.
In particular, packets that use IP at the network layer, TCP at the transport layer, and \gls{iec104} at the application layer.

To distinguish between illegitimate and legitimate traffic, the \gls{sids} uses a set of rules defined in a machine-readable input file derived from the \gls{sb} (cf. Section~\nameref{subsec:framework_specification}).
Here, specifications are defined as sets of information that represent the known parameters and characteristics of the \gls{sg} infrastructure to some extent.
Anything specified in \gls{sb} is considered valid;
anything that does not conform to a specification is considered malicious traffic.

When observing network traffic, the \gls{sids} examines each packet using \gls{dpi} (cf. Section~\nameref{subsec:framework_deeppacketinspect}).
In this context, the conformance of the data packet to the \gls{sb}, such as the protocols used, protocol fields, address validation, and payload consistency is checked.
After the initial inspection of the packet, the next inspection step evaluates connection attributes and states.
Each connection is defined in \gls{sb}, specifying the properties of the connection, and two Mealy automata modeling the connection endpoints.

Regarding \gls{iec104} traffic within the \gls{scada} network, the roles of endpoints are represented as \gls{mtu} and \gls{rtu}.
For correct semantic mapping of packets, the mapper component is responsible for translating the packet contents into an input symbol for the Mealy automata (cf. Section~\nameref{subsec:framework_automata}).
After the mapper receives the corresponding input symbol, the connection object passes it to the instantiated \glspl{am} representing the connections.
Because of the use of Mealy automata, it provides immediate feedback by returning an output symbol.
If the output symbol indicates an error or suspicious behavior, the connection object triggers the alert generator with a specific alert reason to issue an alert (cf. Section~\nameref{subsec:framework_alert}).
Alert generation is triggered by various components for different reasons.
The cyber threat information database represents the collection of alerts combined with the specification of the infrastructure, which is part of a higher-level correlation~\anc{ as presented in our previous work~\cite{24_sen2021towards,sen2022contextual}}.

\subsection*{Specification Basis} \label{subsec:framework_specification}
Based on a formal \gls{gim} of \gls{sg}~\cite{4_klaer_sgam_2020}, we extract the \gls{sb} for the \gls{sids} through the explicit data model definition.
Thus, whitelists can be created from the data model and anomaly detection through the whitelist configurations.
This includes communication (e.g., link quality, routing, packet flows), authentication (e.g., MAC/IP addresses), and process data (e.g., control, measurement, state - plausibility).
Table~\ref{tab:framework_specification} describes the domains and information fields of the \gls{sb}.

\begin{table}[h!]
\centering\small
\caption{Domain-specific attribution of captured traffic.}
\begin{tabular}{||p{2.3cm} p{9cm}||}
 \hline
 Domain & Field Attribution\\
 \hline\hline
 Communication                              & Address matching of packets (L2-L4, L7).\\
                                            & Connection and established communication channel (client/server, protocol, port).\\
                                            & Packet flow according to protocol (L4, L7).\\
 Asset                                      & Data point matching.\\
                                            & Integrity at data point level.\\
                                            & Role-based verified operations.\\
 Operation                                  & Technical assets boundaries.\\
                                            & Technical command execution capability.\\
 \hline
\end{tabular}
\label{tab:framework_specification}
\end{table}

The \gls{sb} provides information on the behavior of communication, assets, and operating limits, from which rules can be derived.
In the area of communication, e.g., the addresses of the relevant fields of the protocol layer (L2-L4, L7), such as the MAC address, the IP address, the Port number, and \gls{ioa} of the \gls{iec104} protocol, are specified.
Additionally, legitimate connection channels and routes are defined which specify allowed communication channels between the endpoints with protocol types and Port numbers.
In the dynamic scope, the standards assigned to the application layers (L7) are defined accordingly, which then sets the corresponding predefined \glspl{am} for attack detection.
E.g., for \gls{iec104}, \glspl{am} are used that represent the data transaction process during communication initialization and confirmation of control commands.

The use of protocols is also considered in the \gls{sb}.
E.g., the use of certain protocols such as SSH can either be whitelisted or even restricted to certain periods such as maintenance times on weekends.
Protocol behavior is observed with \glspl{am} that represent valid communication flows for specific protocols.
Currently, there is only one model for \gls{iec104} traffic, but in general other state-full protocols can also be modeled through \glspl{am}.
The \gls{sb} can also be extended to include other criteria such as the maximum \gls{rtt} for TCP packets.

In the context of resource behavior, data points are taken from \gls{sb} and verified for legitimacy within industrial telecontrol protocols, that is, regarding known data points with correct addressing.
Consequently, the integrity of the data points is defined according to \gls{sb}, if the data characteristics within the data points (e.g. \gls{ioa} in \gls{iec104}) are correctly assigned to the asset in the right communication direction.
This provides the base for role-based verification of asset operations, in which the legitimacy of operation options of assets is also defined by the data points (e.g., sensors can send measurements but not commands).

In the scope of operational behavior, the technical operating boundaries of assets (e.g., maximum power rating for setpoints), and the execution plausibility of commands are also extracted from the domain-specific knowledge of the \gls{sb} (e.g., nominal power-dependent plausibility range for the $\cos\phi$ setting of inverters~\cite{scheben2017status}).

\subsection*{Deep Packet Inspection} \label{subsec:framework_deeppacketinspect}
The functionality \gls{dpi} is a key feature of \gls{sids} and is anchored in the central organizer and forwarder of all internal intrusion detection processes (cf. Figure~\ref{fig:framework_overview}).
After a packet is received, it is categorized depending on the packet layers it contains.
Relevant packets are those that correspond to one of the protocols described in the \gls{sb}, e.g., \gls{iec104} or SSH packets.

The categorization determines the checks that are performed on each packet.
Packets that are classified as irrelevant are ignored, while the contents of packets containing industrial protocols such as \gls{iec104} are checked more thoroughly.
Although basic address detection and verification are performed at the first level of \gls{dpi}, advanced and contextual checks are performed as part of connection-related checks.
Each packet associated with a particular connection object is forwarded to the corresponding checks.
Connection objects represent a connection between two endpoints.
Each of the endpoints is assigned specific addresses for each network layer, including the application layer, e.g., the \gls{iec104} protocol, for which an \gls{am} is assigned.
For \gls{iec104} traffic, this means that each connection contains two \glspl{am}, an \gls{mtu} model, and an \gls{rtu} model.

When a packet is assigned to a connection by the \gls{dpi} component, all addresses are checked for consistency, both on the sender and the receiver side.
In addition, the flow control of the \gls{iec104} layer is also checked.
For this task, connections store the current packet sequence control counters for each endpoint individually.
When a connection receives a packet containing sequence control information, namely packets containing \gls{apci} frames in I\hyp{}frame and U\hyp{}frame format, the connection objects are checked.
They are checked for both endpoints whether the sequence numbers match the current counters and transmission direction.
In addition, an I\hyp{}framed \gls{apci} indicating an \gls{asdu} is checked for technical specification conformity.

After all addresses, traffic sequence, and technical specifications are checked, the packet is passed to the packet mapper.
The mapper maps the packet to an input symbol of the automata alphabet as an automata input.
If any of the checks of \gls{dpi} fails, e.g., the contained address information is unknown or the packet cannot be assigned to a connection, an alert is issued.

\begin{figure}
    \centerline{\includegraphics[width=\columnwidth]{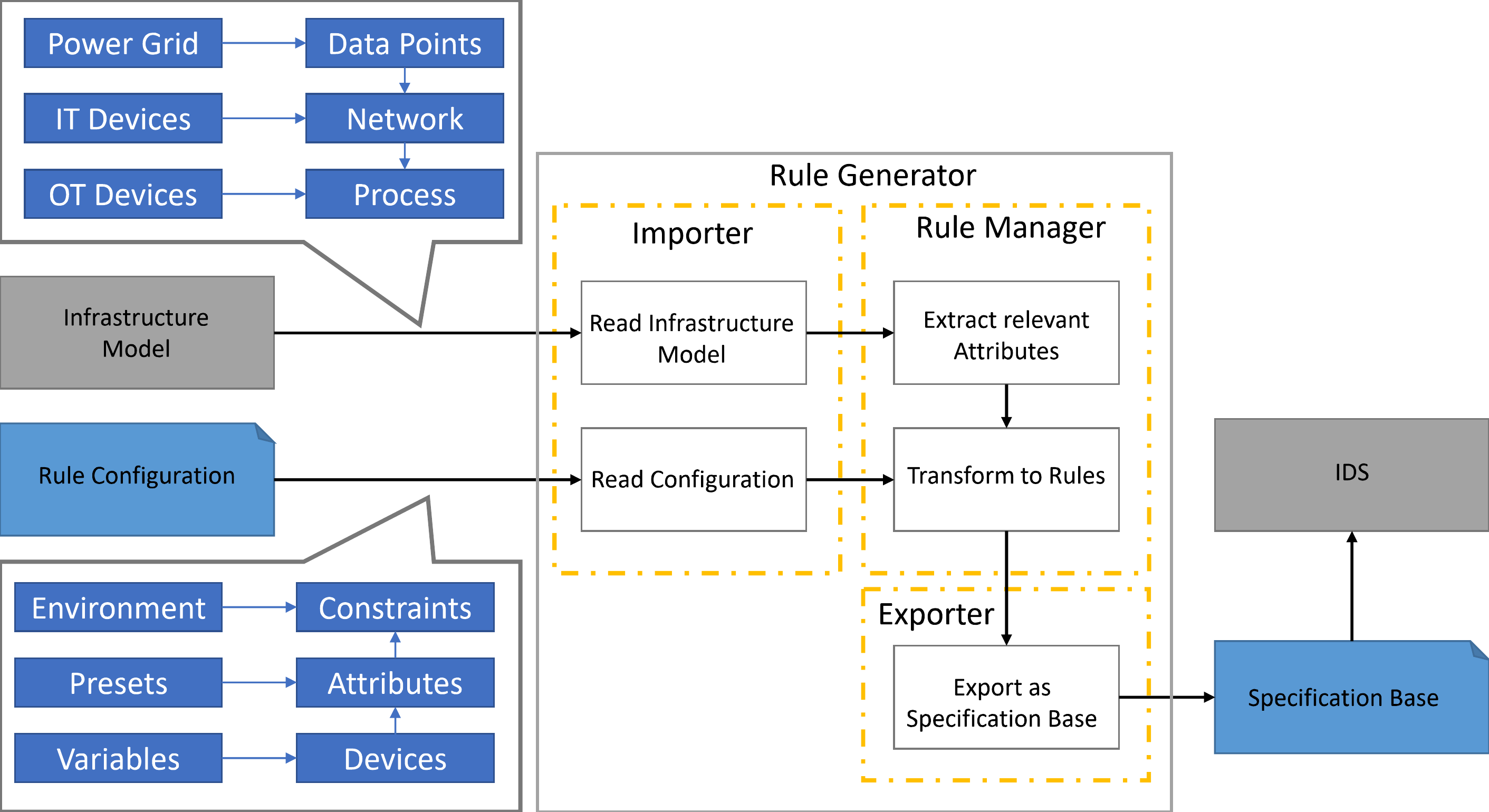}}
    \caption{Illustration of the rule generation process to automatically generate the \gls{sb} based on the infrastructure model.}
    \label{fig:framework_rulegenerator}
\end{figure}

The automated process of generating the \gls{sb} based on the infrastructure model is broken down into several components as shown in Figure~\ref{fig:framework_rulegenerator}.
The input consists of an \gls{gim}, which describes the infrastructure in the respective domains of power grid, IT, and \gls{ot} devices.
Each of these domains contains domain-specific information, such as asset data points, component networking, and their operational function in the process.
In addition, a configuration is required that specifies the rules that will be used later to detect attacks.
The rule configuration specifies the type of devices of interest for which rules are to be created that contain attributes and their environmental constraints.

To achieve the desired detection quality, the \gls{sids} must be correctly configured by the given input.
The prerequisite for this is \gls{sb}, which is to be generated by the rule generator.
The task of the rule generator is to convert a \gls{gim} into a \gls{sb} based on a given configuration.
This \gls{sb} represents the set of rules that the \gls{sids} uses to decide which communication and payload content is valid.

The rule generator consists of three modules, each serving a different purpose.
The importer is used to read the respective inputs - the \gls{gim} and the configuration - and prepare them for further use.
The rule manager is the main module of the rule generator and is responsible for reading the relevant attributes from the \gls{gim} and converting them into rules based on the specifications in the configuration.
Finally, the exporter summarizes the generated rules in a \gls{sb} that can be read by the \gls{sids}.
After generating the \gls{sb}, the \gls{sids} can apply the previously generated specifications to the packets of the captured network communication.
As soon as the given specifications are violated or the recorded communication deviates from the expected normal behavior, alarms are triggered.

\subsection*{Automata Model} \label{subsec:framework_automata}
\begin{figure}
    \centerline{\includegraphics[width=\columnwidth]{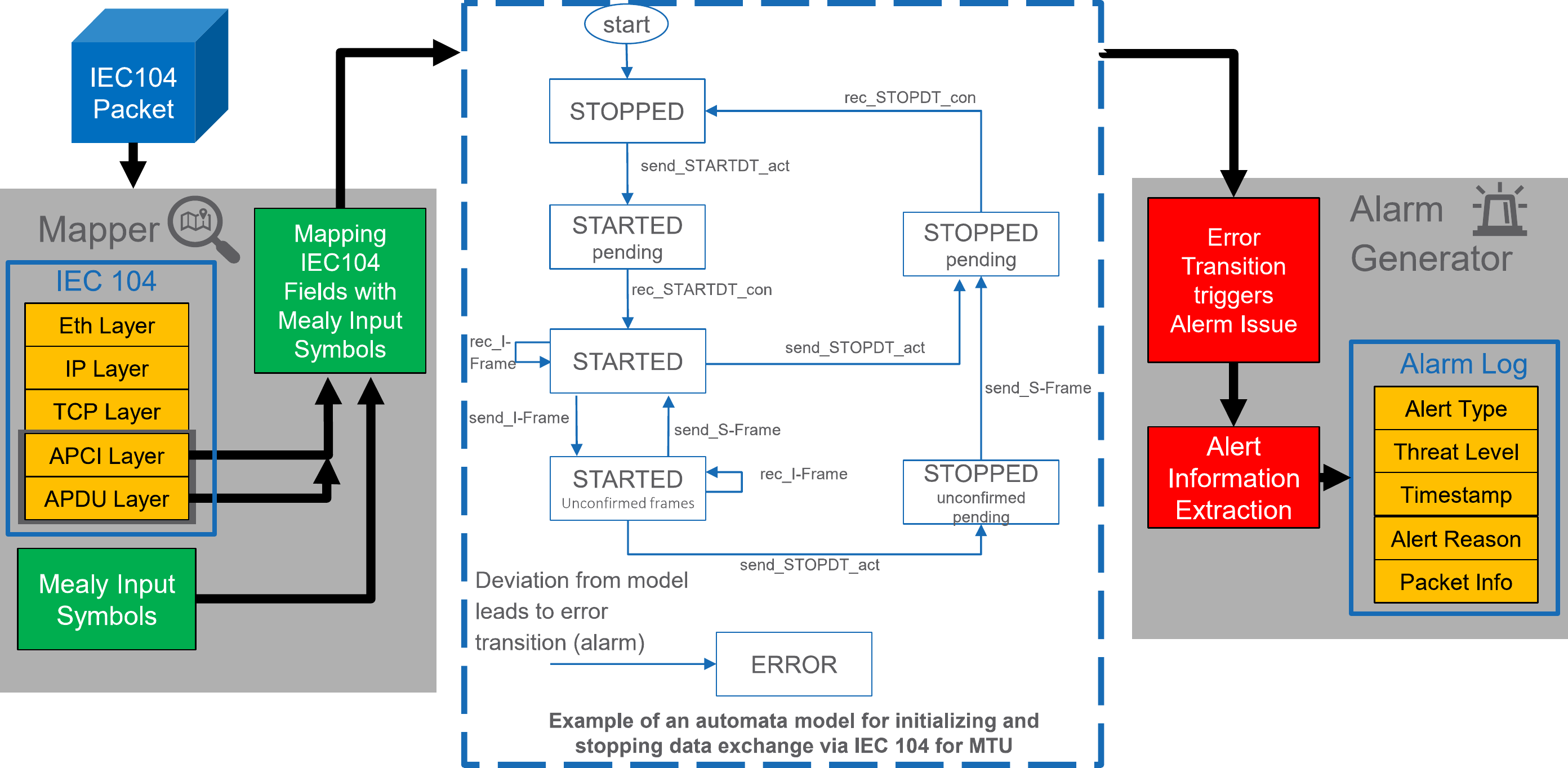}}
    \caption{Exemplary illustration of the mapper, that maps \gls{iec104} data packets to input symbols for the automata, which here, in an example for the \gls{mtu} endpoint connection, generates according to he protocol standard a Mealy \gls{am} to stateful monitor e.g., the data transaction.}
    \label{fig:framework_automata}
\end{figure}
\glspl{am} are used to dynamically check multiple packets within industrial protocol traffic (cf. Figure~\ref{fig:framework_automata}).
The goal of \glspl{am} is to model flow-based processes within the communication process according to the selected protocol, such as \gls{iec104}.
The states of the \gls{am} represent, e.g., the start of a connection and data transmission, the tracking of pending acknowledgments of commands and measured values, and the stop of the data transmission.
Packets identified as industrial protocol traffic are processed by the mapper component, which maps the packets to their corresponding input alphabet counterparts for the \gls{am}.

The mapper translates the contents of the \gls{iec104} layers into input symbols for the \gls{mtu} and \gls{rtu} automata used within the connection object.
A data packet may contain multiple instances of \gls{iec104} layers.
The mapper then returns an ordered list of symbols mapped from these packets.
The symbol is mapped based on the contents of the decoded \gls{iec104} layer.
Within this process, the format of the observed \gls{apci} frame is determined, where the \gls{apci} frame can have three different formats: U\hyp{}frame, I\hyp{}frame, and S\hyp{}frame.
Depending on the determined format, the frames are checked for additional flags that indicate membership in specific groups of packets mapped to special input symbols.
Subsequently, these input symbols represent the set of possible input symbols for the \gls{iec104} automata.
Additionally, an error symbol is used to indicate that the packet does not match any of the criteria used to assign it to one of the known input symbols.

According to the \gls{iec104} protocol standard, two transition systems are modeled, one representing the \gls{mtu} stations of \gls{scada} networks and one modeling \gls{rtu} station for each connection.
The use of this role-based modeling approach allows the states within the \glspl{am} to be sufficiently differentiated, such as the states of connection establishment and valid data transmission.
To define the processes within the connection procedure of communication more accurately, the models must also be able to determine whether an input is sent or received.
To this end, each packet categorization is extended to include a prefix indicating whether the packet was sent or received.
Both automata use the same input alphabet and state sets, but differ slightly in some transitions.

After receiving an input symbol and using a transition, the automata returns an output indicating whether the input results in a suspicious state or whether this packet type is invalid for the current protocol procedure.
Internally, this is done by a status variable within the automaton object.
When certain transitions are used, they trigger functions that change the internal state, which is always given as a return value after processing the input.
E.g., the \gls{am} requires the generation of 15 different input symbols for seven different packet types.
The packets are recognized by the mapper and then extended by a prefix indicating the direction of transmission and an error input.
These transitions do not change the internal state variable, so the output would be valid in the sense of Mealy automata.

All other transitions that trigger a change in the internal state variables of \glspl{am} are undefined behavior, i.e., a violation of the protocol procedure.
Therefore, the output for each of these transitions is invalid.
The error input indicates that the packet was not recognized as belonging to one of the defined packet types, therefore cannot be processed, and thus leads to an alert.

\begin{figure}
    \centerline{\includegraphics[width=\columnwidth]{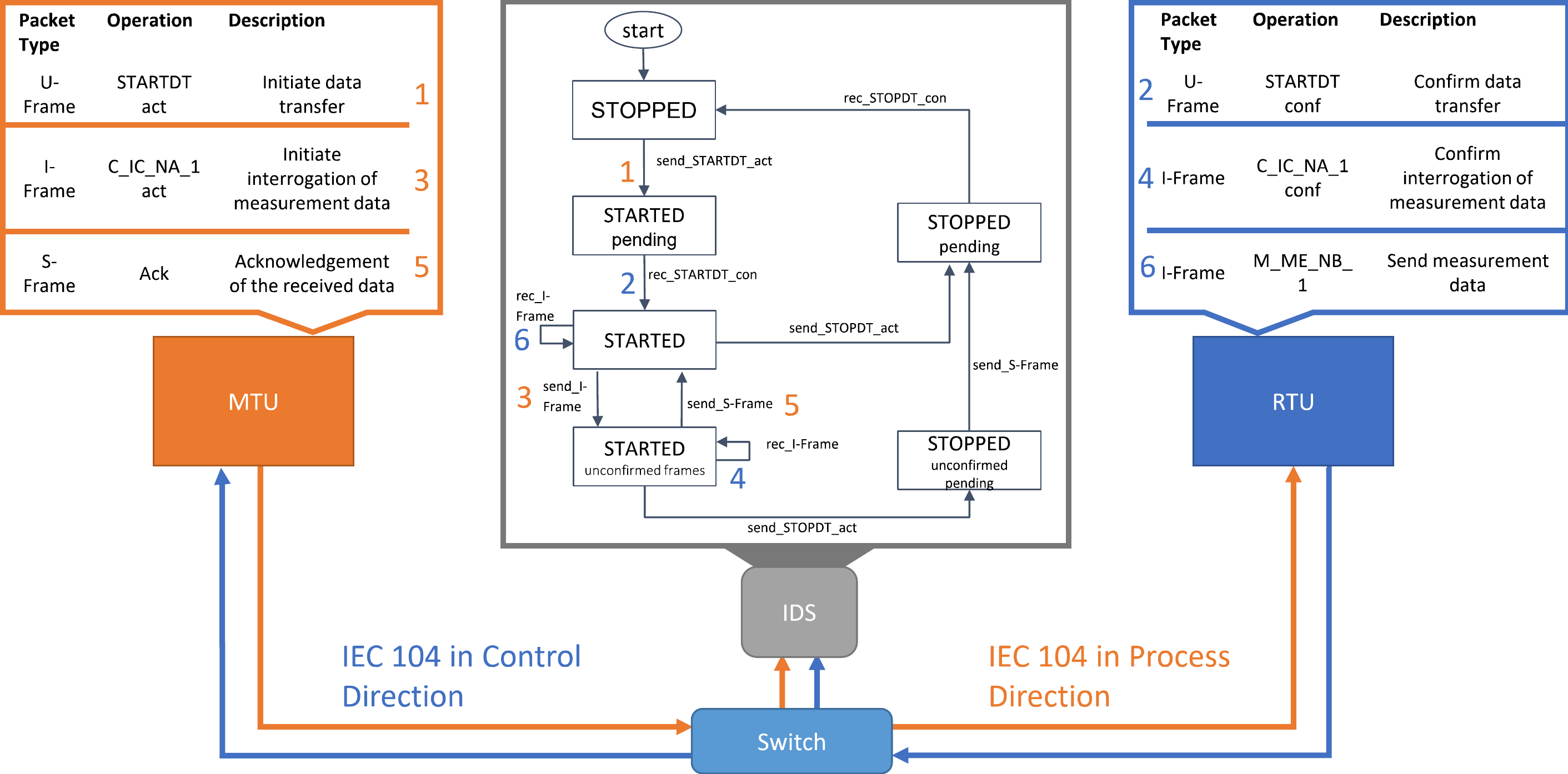}}
    \caption{Example illustration of packet flow conformance checking based on the \gls{mtu} automaton model, showing a simple \gls{mtu} and \gls{rtu} communication scenario.}
    \label{fig:framework_automata_example}
\end{figure}

Figure~\ref{fig:framework_automata_example} illustrates an example of how \glspl{am} works to check the consistency of normal traffic flow in the monitored communication channel.
As an example, a data transmission sequence is used that contains an interrogation operation, where the \gls{mtu} initiates the data transmission.
After starting the data transmission, both automata reach the ``STARTED'' state, which allows I-frames to be sent.
The \gls{mtu} sends an interrogation command to \gls{rtu}, which is acknowledged with the first I-frame back to \gls{mtu}.
After the first I-frame to the \gls{mtu}, the \gls{rtu} sends several I-frames containing the measurement data.
The \gls{mtu} acknowledges the packet reception with an S-frame indicating that all previous frames have been sent correctly and that both automata should now be in the ``STARTED'' state, since both have no unacknowledged frames.
Since the traffic flow conforms to the automaton model, the conformity of the packet flow in this example is therefore also classified as correct by \gls{sids}.

\subsection*{Alert Generation} \label{subsec:framework_alert}
Alerts are the notifications of the \gls{sids} that are triggered when certain packets violate the specification.
They are issued by the alert generator component of \gls{sids} and recorded in a machine-readable log file.

All alerts are written to a log file that assigns a unique running ID to each new alert.
Each alert begins with an ID tag, followed by the attributes specified in Table~\ref{tab:framework_alert}.

\begin{table}[h!]
\centering\small
\caption{Alert output from \gls{sids}}
\begin{tabular}{||p{3.5cm} p{8cm}||}
 \hline
 Alert Field & Description\\
 \hline\hline
 Alert Type                                 & The type of alert indicates what type of alert has occurred.\\
 Threat Level                               & Low, medium or high threat levels.\\
 Timestamp                                  & Each warning issued contains the timestamp when the warning was created.\\
 Alert Reason                               & A textual reason that triggered the creation of this warning message.\\
 Packet Content Information                 & Detailed information about which data packet content is related to the issued alert.\\
 \hline
\end{tabular}
\label{tab:framework_alert}
\end{table}

To illustrate how alert messages are generated, we provide an example in Listing~\ref{lst:alert_file}.
We use Metasploit's \gls{iec104} Client Utility Module as the basis for this sample scenario~\cite{29_metasploitiec104}.
Therefore, the scenario underlying this example is that a new endpoint with unknown IP and MAC address acts as a \gls{mtu} and attempts to establish a \gls{iec104} connection to a \gls{rtu}.
In doing so, the new \gls{mtu} also sends a control command specifying a new setpoint, such as a new power injection for a \gls{pv} inverter.
In this example, the generated alarms cause anomalies regarding the IP and MAC addresses of the new endpoint, as these are not specified in the \gls{sb}.
In addition, the connection is also not valid because the communication channel between the new endpoint and the \gls{rtu} is also not specified.
Thus, any commands sent from the new endpoint to the \gls{rtu} are also considered invalid.
Furthermore, the control command contained a setpoint that also violates the specified allowable range of valid setpoints.
Thus, all active interactions between the new endpoint and the \gls{rtu} are classified as anomalies and output as alerts.

\lstset{language=XML, numbers=none, captionpos=b, xleftmargin=0.2em, caption=Example of alert messages generated by the alert generator. , label=lst:alert_file, frame=single , breaklines=true, basicstyle=\scriptsize, multicols=2}
\begin{lstlisting}
[ALERT_0]
alert_type = IP_MISMATCH
threat_level = high
timestamp = 14.04.2022 10:47:09
alert_reason = IP of this packet is unknown: 173.24.0.3
packet_info = ETH / IP / TCP / IEC104-U

[ALERT_1]
alert_type = PORT_MISMATCH
threat_level = high
timestamp = 14.04.2022 10:47:09
alert_reason = One of the Ports of this packet is unknown: 59478
packet_info = ETH / IP / TCP / IEC104-U

[ALERT_2]
alert_type = NO_SUCH_CONNECTION
threat_level = high
timestamp = 14.04.2022 10:47:09
alert_reason = Connection does not exist in whitelisting data!
packet_info = ETH / IP / TCP / IEC104-U

[ALERT_3]
alert_type = INVALID_OPERATION
threat_level = high
timestamp = 14.04.2022 10:48:00
alert_reason = Send packet contains invalid operation for the endpoint!
packet_info = ETH / IP / TCP / IEC104-I

[ALERT_4]
alert_type = INVALID_SETPOINT
threat_level = high
timestamp = 14.04.2022 10:48:00
alert_reason = Active control command contains invalid setpoint!
packet_info = ETH / IP / TCP / IEC104-I
\end{lstlisting}

Overall, our \gls{sids}, which automatically derives its \gls{sb} based on a formal \gls{gim}, is designed to detect explainable anomalies from different domains.
Specifically, the domain-specific knowledge used for anomaly detection is extracted as appropriate rules from the \gls{gim}, which represents the operator's existing knowledge of its infrastructure, without requiring the knowledge of cyber-security experts.
Moreover, the dynamic nature of communication behavior is also validated by \glspl{am} with respect to protocol conformance and flow consistency.
Subsequently, both the dynamic packet flow and attributes, as well as the protocol field values within the payload, are validated and checked for potential inconsistencies or specification violations.
Thus, our \gls{sids} detects critical violations of legitimate processes and infrastructure specifications at both the communication and operational levels, relying only on existing and available knowledge without requiring external expertise.

\section*{Evaluation \& Discussion} \label{sec:result}
To demonstrate and discuss the performance of our proposed \gls{sids}, we evaluate its detection quality in a physical testbed for attack and non-attack scenarios.

\subsection*{Smart Grid Testbed Setup} \label{subsec:result_testbed}
\begin{figure}
    \centerline{\includegraphics[width=\columnwidth]{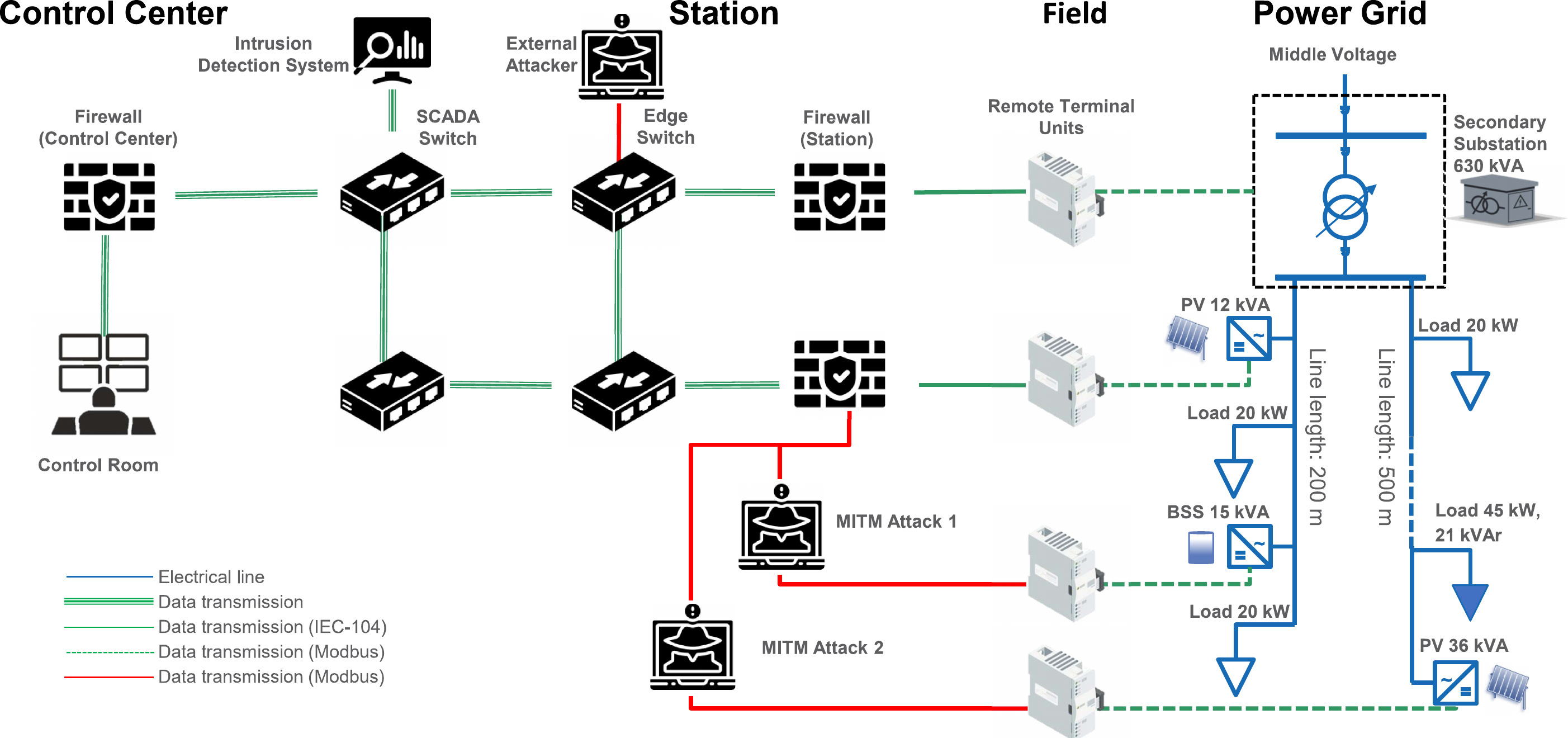}}
    \caption{Our setup of a \gls{sg} testbed
    replicates a distribution grid with \glspl{der} and \gls{ict} infrastructure for control and monitoring.}
    \label{fig:tesbed_setup}
\end{figure}

We evaluate \gls{sids} in a cyber-physical testbed as shown in Figure~\ref{fig:tesbed_setup}\anc{ that is based on our previous work~\cite{25_sen2021towards}}.
The testbed replicates an MV / LV grid consisting of physical components networked through a dedicated \gls{ict} infrastructure.
Therefore, neither simulations nor virtualized components are involved.
Since our \gls{sids} is designed to monitor specific communication channels within defined network segments and thus acts more like a sensor with more than one entity deployment, the limited complexity of the testbed does not limit the scope of the study.

In our test setup, we use electrical equipment such as a 640 kVA secondary substation, 22 kWh \gls{bss}, 12 kVA and 36 kVA \glspl{pv}, and several resistive/inductive loads.
The power system topology consists of two strings to which the \glspl{der} and loads are connected, and on which we can measure current and voltage via integrated three-phase measuring points.
We control the \gls{der} through their Modbus interface via \glspl{rtu}, which is provided by their respective inverter.
Following a \gls{scada} network, the testbed also includes a process network consisting of the \gls{ict} infrastructure and the control room.
The control room represents an \gls{mtu} that sends \gls{iec104} control and query commands to \glspl{rtu}.

We consider different attack scenarios based on attacks that have already gained access to the process network.
The external attacker represents a new entity within the system with unknown IP and MAC addresses.
Contrary, the \gls{mitm} attack intercepts the communication between the control room and the selected \glspl{rtu}.

In addition, our \gls{sids} approach is deployed in the process network at the \gls{scada} switch, where traffic is monitored via a mirrored SPAN port or dedicated network taps.
Active inline network taps are used between the \gls{scada} node and switches to capture \gls{scada} traffic and perform timestamping with a high resolution of 8 ns~\cite{profishark2018}.
The mirrored SPAN port allows all traffic passing through the target switch to be captured, but with lower timestamp resolution and possibly more jitter.
Preferably, dedicated network taps are used for the main communication channel between the \gls{mtu} and \gls{rtu} to be monitored.
Thus, \gls{scada}-related network traffic is continuously forwarded to \gls{sids} for intrusion detection.

\subsection*{Methodology} \label{subsec:result_proecdure}
We use the cyber-physical testbed and attack scenario described in~\nameref{subsec:result_testbed}.
In the first scenario without attack induction, all stations and traffic are within the specification, where normal protocol behavior of a system operation under normal conditions is replicated (Scenario 1).
We use this scenario to examine the \gls{sids} under normal operation conditions with an average traffic volume.

In the second scenario, we investigate the detection quality of \gls{sids} under attack conditions.
To this end, this scenario is divided into replicating an attack from outside the testbed with limited knowledge of the internal network and data points (Scenario 2-a) and an attack with sophisticated knowledge (Scenario 2-b).

Scenario 2-a contains a communication attempt with unknown addresses in L2 and L3 (e.g., MAC and IP address) to \gls{rtu}, sending an interrogation command to query measurement values.
Scenario 2-b is executed as a \gls{mitm}-based \gls{fdi} attack in which two different types of packet are injected.
The first type of injected packet contains an \gls{ioa} that is not associated with the corresponding asset in the specification (Scenario 2-b-I).
In contrast, the second type contains an \gls{ioa} included in the specification and also regularly used in the normal conversation of \gls{mtu} and \gls{rtu} with the correct mapping of devices (Scenario 2-b-II).
However, the measurements transmitted with these packets contain measurements that are overlaid with small noise within the range of the technical specification.

We also measure the performance of \gls{sids} in large traffic volumes per connection in a short time.
An important requirement for this investigation is that \gls{sids} always observes the beginning of the connection to use \glspl{am} for correct context and packet sequence tracking.
Otherwise, it cannot find the correct initial state and therefore produces alerts for almost all data packets.

To evaluate the quality of detection, we rely on the confusion matrix and the performance metrics derived~\cite{28_tharwat2020classification}.
Thus, we measure the following metrics:
\begin{itemize}
 \item \gls{tp}: event correctly classified as attack indicator
 \item \gls{fp}: event incorrectly classified as attack indicator
 \item \gls{tn}: event correctly not classified as attack indicator
 \item \gls{fn}: event incorrectly not classified as attack indicator
\end{itemize}
To evaluate the performance of our approach, we also use the following telemetry data of the captured network traffic in the scenarios.
\begin{itemize}
 \item L2: MAC source and destination addresses
 \item L3: IP source and destination addresses, checksum
 \item L4: TCP sequence and acknowledge numbers, port number, checksum, \gls{rtt}
 \item L7: IEC 104 protocol fields of U- and I-frames
\end{itemize}

\subsection*{Evaluation of Performance \& Classification Accuracy} \label{subsec:result_res}
\begin{table}[h!]
\centering\small
\caption{Confusion Table of Experiments.}
\begin{tabular}{||p{2.3cm} p{1cm} p{1cm} p{1cm} p{1cm}||}
 \hline
 Scenarios & \gls{tp} & \gls{tn} & \gls{fp} & \gls{fn}\\
 \hline\hline
 1                                          & 0 & 200 & 0 & 0\\
 2-a                                        & 115 & 0 & 0 & 0\\
 2-b-I                                      & 10 & 0 & 0 & 0\\
 2-b-II                                     & 0 & 0 & 0 & 10\\
 \hline
\end{tabular}
\label{tab:result}
\end{table}
Our results are presented in Table~\ref{tab:result}.
Within Scenario 1, our results indicate that \gls{sids} does not generate alerts caused by addressing, automata errors, or sequence number violations.
The only parameter that may cause slight variations in the \gls{fpr} is the maximum \gls{rtt} parameter, which in our experiments was parameterized in the range of $150 ms$ and $200 ms$.
Narrower ranges caused more \glspl{fp} in our experiments due to varying \gls{rtt} in the communication channels.
With this adjustment, using a sufficiently large value for the maximum \gls{rtt} (e.g., upper $95\%$ confidence interval of the \gls{rtt} variance), no \glspl{fp} were produced.

In Scenario 2-a, the attacker mimics the normal behavior of a \gls{mtu} by starting a conversation and sending a query command for the measurement data.
The \gls{rtu} responds with measurement data.
All $115$ malicious packets were correctly detected.

Within Scenario 2-b, we inject a total of $20$ packets (10 packets from each of the sub-scenarios).
\gls{sids} was able to correctly classify the $10$ packets from scenario 2-b-I (\gls{tp}) due to incorrect addressing of \gls{ioa}.
Scenario 2-b-II represents an edge case where the attacker performs perfect spoofing and adheres to the legitimate specification of the system.
Therefore, the $10$ injected packets from Scenario 2-b-II were not correctly detected by \gls{sids}, showing the limits of our approach (\gls{fn}).
However, limiting the range of attack actions so that the attacker can evade detection can shift the impact trajectory of the attack into a treatable scope.
Subsequently, a larger scope of attack is required to cause more impact, imposing more actions on the attacker that can potentially reduce their stealthy movement.

To assess processing performance, we also evaluated the processing time of packets with and without specification compliance (Scenario 1 and Scenario 2).
For compliance with the specification (Scenario 1), each packet monitored by \gls{sids} is processed on average at $0.3ms$ with an insignificant standard deviation.
With invalid traffic (Scenario 2), each packet is processed in an average of $1.5ms$ with also insignificant standard deviation.
The reason for this discrepancy is that when a packet violates the specification, several steps are triggered in the reporting mechanism to extract alert-relevant information, which is written to the alert log.

In addition, we have also performed a comparison with other intrusion detection approaches, which is shown in Table~\ref{tab:evalation_comparision}.
However, due to the lack of a standardized benchmark evaluation for countermeasures against cyber-attacks in \gls{scada} systems, the comparison is qualitative.
\begin{table}[h!]
\centering\small
\caption{Comparison with other intrusion detection approaches.}
\begin{tabular}{||p{1cm} p{1.5cm} p{1cm} p{1.5cm} p{1cm} p{0.75cm} p{0.75cm} p{0.75cm} p{0.75cm}||}
 \hline
 Ref. & Tech. & Proto. & Environ. & Att. & Mat. & Exp. & Det. & Perf.\\
 \hline\hline
 ~\cite{41_al2016oscids} & ontology & Modbus TCP & simulation & mixed & M & L & M & L \\
 ~\cite{43_cruz2016cybersecurity} & machine-learning & Modbus TCP & testbed & protocol & H & H & M & H \\
 ~\cite{46_udd2016exploiting} & specification & \gls{iec104} & simulation & channel & H & L & M & H  \\
 ~\cite{51_yang2016multidimensional} & specification & IEC 61850 & testbed & packet & H & M & L & H\\
 ~\cite{52_adepu2018distributed} & automata & Eth / IP & testbed & location & H & H & L & N/A\\
 ~\cite{55_lin2016runtime} & semantic & DNP3 & simulation & control & H & L & M & L \\
 ~\cite{57_wang2018intrusion} & time series & \gls{iec104} & simulation & \gls{scada} & L & L & M & N/A\\
 our \gls{sids} & specification & \gls{iec104} & testbed & \gls{fdi} & H & H & H & M \\
  \hline
\end{tabular}
\label{tab:evalation_comparision}
\end{table}
The comparison compares our \gls{sids} qualitatively with other approaches based on the following metrics:
\begin{itemize}
 \item Tech.: describes the detection basis of the methodology
 \item Proto.: describes which protocol is the main target of protection
 \item Environ.: describes whether a simulated or physical testbed was used
 \item Att.: describes on which basis the attack scenarios were designed
 \item Mat.: describes the degree of readiness of the approach in likert-scale
 \item Exp.: describes the flexibility to be adapted to other protocols in likert-scale
 \item Det.: describes the degree of detection quality of the approach in likert-scale
 \item Perf.: describes the performance level of the approach in likert-scale
\end{itemize}
As the comparison shows, the conditions and environment under which the different approaches were evaluated are mostly different.
The attack scenarios also diverge in their scope, vectors used, and interaction with operational equipment.
The experiments conducted also differ within their respective environments where simulation was used with simplification and abstraction.
Many of the approaches have a high degree of maturity and are capable of being deployed and operated in real grid environments.
However, they lack the ability to be extended to other protocols.
The performance of the approaches shows the recognition capabilities of packets within the time span $0.1ms$ to $1s$, and the detection quality is also in the medium range, which is mainly due to the high \gls{fn}.
Thus, the evaluation suggests that our \gls{sids} enables reliable detection of cyber-attacks within a reasonable time.

\subsection*{Discussion} \label{subsec:result_dis}
The results show that for normal operation, our \gls{sids} has not triggered any (false) alert messages.
Deviations were only caused by a too narrow \gls{rtt} range and should be considered when carefully setting this parameter for detection quality.

In the attack-induced scenario (Scenario 2-b-II), where the attacker knows which addresses and \gls{ioa} entries are valid, \gls{sids} will have difficulty detecting them if the manipulated values are within the technical specification.
However, any deviation from the addresses defined in the specifications will lead to high detection rates.
In general, the detection quality is very dependent on the provided \gls{sb}.
The given structure of the \gls{sb}, which defines the exact addressing for each allowed connection, are very strict rules that detect all connections that are not explicitly allowed.
Attacks from outside with limited knowledge of the technical specifications of infrastructures can thus be reliably detected.

To create perfect spoofing conditions, the attacker must maintain complete consistency and compliance with the specification, which requires extensive knowledge.
Furthermore, the attacker must perform prior steps, such as reconnaissance and lateral movement, to persist in the process network, potentially leaving traces in the communication layer.
In the context of situational awareness for intrusion detection, our \gls{sids} can act as a low-level sensor that provides domain-specific indicators of multi-staged cyber-attacks.
Alerts can be centrally processed with other indicators from other \gls{ids} sensors through a correlation system based on \gls{siem} to reconstruct the attack sequence~\cite{24_sen2021towards,sen2022contextual}.

While our evaluation focuses on \gls{iec104}, the proposed \gls{sids} can also be used for other \gls{scada} protocols such as IEC-61850.
The semantics of \gls{sb} is provided by \gls{gim}, where the adaptation of a new protocol requires the mapping process of the data and the fields of the protocol.
Thus, an appropriate mapper must be developed to reference semantic data with protocol fields.
In addition, \gls{am} can generally be adapted to stateful communication such as TCP-based protocols, where packet flows can be described with state transitions.

\section*{Conclusion} \label{sec:conclusion}
In the context of power grids transitioning to \glspl{sg}, countermeasures against sophisticated cyber-attacks based on reliable detection mechanisms are required.
To this end, we present a \gls{sids} that uses a graph-based specification to holistically encapsulate the \gls{sg} infrastructure to detect cyber-attacks.
We discuss the design and subsequent implementation of our \gls{sids}, which consists of a \gls{dpi} component and an \gls{am}.
Using our implementation, we evaluated the detection quality within a physical testbed for different scenarios under attack and normal conditions.

Our main findings are that our \gls{sids} approach can reliably detect attackers injecting false data into intercepted \gls{iec104} channels.
The performance and detection quality show the advantages of an approach \gls{sids} and was validated in our study.
Moreover, the disadvantage of high knowledge provisioning overhead is reduced by our novel approach of coupling infrastructure modeling with \gls{sids}.
Future work includes investigating different methods for detecting \gls{fdi} in a cooperative, neighborhood-oriented manner.
In addition, the generated alerts of the proposed \gls{sids} will also be investigated in terms of providing a reliable basis for a higher-level correlation system for reconstructing complex attack campaigns.


\begin{backmatter}

\printglossary

  
\section*{Funding} 
\anon{This work has partly been funded by the German Federal Ministry for Economic Affairs and Climate Action (BMWK) under project funding reference 0350028.}

\section*{Availability of data and materials} 
  No data and materials are published.

\section*{Author's contributions}
\anon{
 \begin{inparaenum}[i)]
  \item  Conceptualization: Ömer Sen;
  \item  Methodology: Ömer Sen, Maik Lühmann, Florian Sprünken;
  \item  Validation:  Ömer Sen, Maik Lühmann, Florian Sprünken, Martin Henze;
  \item  Formal analysis: Maik Lühmann, Florian Sprünken, Ömer Sen;
  \item  Investigation: Maik Lühmann, Florian Sprünken, Ömer Sen, Martin Henze;
  \item  Resources: Maik Lühmann, Florian Sprünken, Ömer Sen;
  \item  Data Curation: Ömer Sen, Maik Lühmann, Florian Sprünken;
  \item  Writing - Original Draft: Ömer Sen;
  \item  Writing - Review \& Editing: Ömer Sen, Maik Lühmann, Florian Sprünken, Martin Henze, Dennis van der Velde, Immanuel Hacker, Andreas Ulbig;
  \item  Visualization: Ömer Sen, Immanuel Hacker;
  \item  Supervision: Ömer Sen, Martin Henze, Andreas Ulbig;
  \item  Project administration: Michael Andres, Dennis van der Velde;
  \item  Funding acquisition: Michael Andres;
 \end{inparaenum}
}

\section*{Competing interests}
  The authors declare that they have no competing interests.

\bibliographystyle{bmc-mathphys} 


\newcommand{\BMCxmlcomment}[1]{}

\BMCxmlcomment{

<refgrp>

<bibl id="B1">
  <title><p>{Review on smart grid control and reliability in presence of
  renewable energies: Challenges and prospects}</p></title>
  <aug>
    <au><snm>Ourahou</snm><fnm>M</fnm></au>
    <au><snm>Ayrir</snm><fnm>W</fnm></au>
    <au><snm>Hassouni</snm><fnm>B EL</fnm></au>
    <au><snm>Haddi</snm><fnm>A</fnm></au>
  </aug>
  <source>Mathematics and computers in simulation</source>
  <publisher>Elsevier</publisher>
  <pubdate>2020</pubdate>
</bibl>

<bibl id="B2">
  <title><p>SMART GRIDS: Fundamentals and Technologies in Electric Power
  Systems of the Future</p></title>
  <aug>
    <au><snm>BERND</snm><fnm>M</fnm></au>
    <au><snm>BUCHHOLZ</snm><fnm>STYCZYNSKI</fnm></au>
    <au><snm>ZBIGNIEW</snm><fnm>A</fnm></au>
  </aug>
  <publisher>Axel-Springer-Strasse, Berlin: SPRINGER-VERLAG BERLIN
  AN</publisher>
  <pubdate>2021</pubdate>
</bibl>

<bibl id="B3">
  <title><p>{Methods for Actors in the Electric Power System to Prevent, Detect
  and React to ICT Attacks and Failures}</p></title>
  <aug>
    <au><snm>Velde</snm><fnm>D</fnm></au>
    <au><snm>Henze</snm><fnm>M</fnm></au>
    <au><snm>Kathmann</snm><fnm>P</fnm></au>
    <au><snm>Wassermann</snm><fnm>E</fnm></au>
    <au><snm>Andres</snm><fnm>M</fnm></au>
    <au><snm>Bracht</snm><fnm>D</fnm></au>
    <au><snm>Ernst</snm><fnm>R</fnm></au>
    <au><snm>Hallak</snm><fnm>G</fnm></au>
    <au><snm>Klaer</snm><fnm>B</fnm></au>
    <au><snm>Linnartz</snm><fnm>P</fnm></au>
    <au><cnm>others</cnm></au>
  </aug>
  <source>ENERGYCon</source>
  <pubdate>2020</pubdate>
</bibl>

<bibl id="B4">
  <title><p>{Graph-based Model of Smart Grid Architectures}</p></title>
  <aug>
    <au><snm>Klaer</snm><fnm>B</fnm></au>
    <au><snm>Sen</snm><fnm>{\"O}</fnm></au>
    <au><snm>Velde</snm><fnm>D</fnm></au>
    <au><snm>Hacker</snm><fnm>I</fnm></au>
    <au><snm>Andres</snm><fnm>M</fnm></au>
    <au><snm>Henze</snm><fnm>M</fnm></au>
  </aug>
  <source>SEST</source>
  <pubdate>2020</pubdate>
</bibl>

<bibl id="B5">
  <title><p>{Cyber attack models for smart grid environments}</p></title>
  <aug>
    <au><snm>Eder Neuhauser</snm><fnm>P</fnm></au>
    <au><snm>Zseby</snm><fnm>T</fnm></au>
    <au><snm>Fabini</snm><fnm>J</fnm></au>
    <au><snm>Vormayr</snm><fnm>G</fnm></au>
  </aug>
  <source>SEGAN</source>
  <publisher>Elsevier</publisher>
  <pubdate>2017</pubdate>
</bibl>

<bibl id="B6">
  <title><p>{Analysis of the cyber attack on the Ukrainian power
  grid}</p></title>
  <aug>
    <au><snm>Case</snm><fnm>DU</fnm></au>
  </aug>
  <source>E-ISAC</source>
  <pubdate>2016</pubdate>
</bibl>

<bibl id="B7">
  <title><p>{IEC 62351 Security Standards for the Power System Information
  Infrastructure}</p></title>
  <aug>
    <au><cnm>IEC</cnm></au>
  </aug>
  <pubdate>2016</pubdate>
</bibl>

<bibl id="B8">
  <title><p>{Secure links: Secure-by-design communications in iec 61499
  industrial control applications}</p></title>
  <aug>
    <au><snm>Tanveer</snm><fnm>A</fnm></au>
    <au><snm>Sinha</snm><fnm>R</fnm></au>
    <au><snm>Kuo</snm><fnm>MM</fnm></au>
  </aug>
  <source>IEEE Transactions on Industrial Informatics</source>
  <publisher>IEEE</publisher>
  <pubdate>2020</pubdate>
</bibl>

<bibl id="B9">
  <title><p>{IPAL: Breaking up Silos of Protocol-dependent and Domain-specific
  Industrial Intrusion Detection Systems}</p></title>
  <aug>
    <au><snm>Wolsing</snm><fnm>K</fnm></au>
    <au><snm>Wagner</snm><fnm>E</fnm></au>
    <au><snm>Saillard</snm><fnm>A</fnm></au>
    <au><snm>Henze</snm><fnm>M</fnm></au>
  </aug>
  <source>RAID</source>
  <pubdate>2022</pubdate>
</bibl>

<bibl id="B10">
  <title><p>{Cybersecurity in Power Grids: Challenges and
  Opportunities}</p></title>
  <aug>
    <au><snm>Krause</snm><fnm>T</fnm></au>
    <au><snm>Ernst</snm><fnm>R</fnm></au>
    <au><snm>Klaer</snm><fnm>B</fnm></au>
    <au><snm>Hacker</snm><fnm>I</fnm></au>
    <au><snm>Henze</snm><fnm>M</fnm></au>
  </aug>
  <source>Sensors</source>
  <publisher>Multidisciplinary Digital Publishing Institute</publisher>
  <pubdate>2021</pubdate>
</bibl>

<bibl id="B11">
  <title><p>{A specification-based state replication approach for digital
  twins}</p></title>
  <aug>
    <au><snm>Eckhart</snm><fnm>M</fnm></au>
    <au><snm>Ekelhart</snm><fnm>A</fnm></au>
  </aug>
  <source>CPS-SPC</source>
  <pubdate>2018</pubdate>
</bibl>

<bibl id="B12">
  <title><p>{Role of machine learning algorithms intrusion detection in WSNs: a
  survey}</p></title>
  <aug>
    <au><snm>Baraneetharan</snm><fnm>E</fnm></au>
  </aug>
  <source>Journal of Information Technology</source>
  <pubdate>2020</pubdate>
</bibl>

<bibl id="B13">
  <title><p>{Cybersicherheit der Netzleittechnik: Ergebnisse aus
  Stabilit{\"a}ts-und Sicherheitsaudits}</p></title>
  <aug>
    <au><snm>Mochalski</snm><fnm>K</fnm></au>
  </aug>
  <source>Realisierung Utility 4.0 Band 1</source>
  <publisher>Axel-Springer-Strasse, Berlin: Springer</publisher>
  <pubdate>2020</pubdate>
</bibl>

<bibl id="B14">
  <title><p>{Cybersecurity for Smart Substation}</p></title>
  <aug>
    <au><snm>Hokama</snm><fnm>WS</fnm></au>
    <au><snm>Souza</snm><fnm>JS</fnm></au>
  </aug>
  <source>T\&D LA</source>
</bibl>

<bibl id="B15">
  <title><p>{Machine learning based intrusion detection systems for IoT
  applications}</p></title>
  <aug>
    <au><snm>Verma</snm><fnm>A</fnm></au>
    <au><snm>Ranga</snm><fnm>V</fnm></au>
  </aug>
  <source>Wireless Personal Communications</source>
  <publisher>Springer</publisher>
  <pubdate>2020</pubdate>
</bibl>

<bibl id="B16">
  <title><p>{A False Sense of Security? Revisiting the State of Machine
  Learning-Based Industrial Intrusion Detection}</p></title>
  <aug>
    <au><snm>Kus</snm><fnm>D</fnm></au>
    <au><snm>Wagner</snm><fnm>E</fnm></au>
    <au><snm>Pennekamp</snm><fnm>J</fnm></au>
    <au><snm>Wolsing</snm><fnm>K</fnm></au>
    <au><snm>Fink</snm><fnm>IB</fnm></au>
    <au><snm>Dahlmanns</snm><fnm>M</fnm></au>
    <au><snm>Wehrle</snm><fnm>K</fnm></au>
    <au><snm>Henze</snm><fnm>M</fnm></au>
  </aug>
  <source>CPSS</source>
  <pubdate>2022</pubdate>
</bibl>

<bibl id="B17">
  <title><p>{Towards a Scalable and Flexible Smart Grid Co-Simulation
  Environment to Investigate Communication Infrastructures for Resilient
  Distribution Grid Operation}</p></title>
  <aug>
    <au><snm>Velde</snm><fnm>D</fnm></au>
    <au><snm>Sen</snm><fnm>{\"O}</fnm></au>
    <au><snm>Hacker</snm><fnm>I</fnm></au>
  </aug>
  <source>SEST</source>
  <pubdate>2021</pubdate>
</bibl>

<bibl id="B18">
  <title><p>{The Purdue enterprise reference architecture}</p></title>
  <aug>
    <au><snm>Williams</snm><fnm>TJ</fnm></au>
  </aug>
  <source>Computers in industry</source>
  <publisher>Elsevier</publisher>
  <pubdate>1994</pubdate>
</bibl>

<bibl id="B19">
  <title><p>{Telecontrol equipment and systems—Part 5-104: Transmission
  Protocols—Network Access for IEC 60870-5-101 Using Standard Transport
  Profiles}</p></title>
  <aug>
    <au><cnm>IEC</cnm></au>
  </aug>
  <source>IEC Standard</source>
  <pubdate>2006</pubdate>
</bibl>

<bibl id="B20">
  <title><p>{Modbus Application Protocol Specification V1. 1b3.
  2012}</p></title>
  <aug>
    <au><cnm>MICIE</cnm></au>
  </aug>
  <source>MICIE Consortium</source>
  <pubdate>2020</pubdate>
</bibl>

<bibl id="B21">
  <title><p>{Description and analysis of IEC 104 Protocol}</p></title>
  <aug>
    <au><snm>Matou{\v{s}}ek</snm><fnm>P</fnm></au>
  </aug>
  <source>Faculty of Information Technology, Brno University o Technology,
  Tech. Rep</source>
  <pubdate>2017</pubdate>
</bibl>

<bibl id="B22">
  <title><p>{Man-in-the-middle attack test-bed investigating cyber-security
  vulnerabilities in smart grid SCADA systems}</p></title>
  <aug>
    <au><snm>Yang</snm><fnm>Y</fnm></au>
    <au><snm>McLaughlin</snm><fnm>K</fnm></au>
    <au><snm>Littler</snm><fnm>T</fnm></au>
    <au><snm>Sezer</snm><fnm>S</fnm></au>
    <au><snm>Im</snm><fnm>EG</fnm></au>
    <au><snm>Yao</snm><fnm>ZQ</fnm></au>
    <au><snm>Pranggono</snm><fnm>B</fnm></au>
    <au><snm>Wang</snm><fnm>HF</fnm></au>
  </aug>
  <publisher>IET</publisher>
  <pubdate>2012</pubdate>
</bibl>

<bibl id="B23">
  <title><p>{Power Systems Management and Associated Information Exchange –
  Data and Communications Security – Part 3: Communication Network and System
  Security – Profiles Including TCP/IP}</p></title>
  <aug>
    <au><cnm>IEC</cnm></au>
  </aug>
  <pubdate>2018</pubdate>
</bibl>

<bibl id="B24">
  <title><p>{Securing IEC 60870-5-104 communications following IEC 62351
  standard: lab tests and results}</p></title>
  <aug>
    <au><snm>Todeschini</snm><fnm>MG</fnm></au>
    <au><snm>Dondossola</snm><fnm>G</fnm></au>
  </aug>
  <source>AEIT</source>
  <pubdate>2020</pubdate>
</bibl>

<bibl id="B25">
  <title><p>{Legacy-compliant data authentication for industrial control system
  traffic}</p></title>
  <aug>
    <au><snm>Castellanos</snm><fnm>JH</fnm></au>
    <au><snm>Antonioli</snm><fnm>D</fnm></au>
    <au><snm>Tippenhauer</snm><fnm>NO</fnm></au>
    <au><snm>Ochoa</snm><fnm>M</fnm></au>
  </aug>
  <source>ACNS</source>
  <pubdate>2017</pubdate>
</bibl>

<bibl id="B26">
  <title><p>{A comprehensive survey on network anomaly detection}</p></title>
  <aug>
    <au><snm>Fernandes</snm><fnm>G</fnm></au>
    <au><snm>Rodrigues</snm><fnm>JJ</fnm></au>
    <au><snm>Carvalho</snm><fnm>LF</fnm></au>
    <au><snm>Al Muhtadi</snm><fnm>JF</fnm></au>
    <au><snm>Proen{\c{c}}a</snm><fnm>ML</fnm></au>
  </aug>
  <source>Telecommunication Systems</source>
  <publisher>Springer</publisher>
  <pubdate>2019</pubdate>
</bibl>

<bibl id="B27">
  <title><p>{Intrusion detection and big heterogeneous data: a
  survey}</p></title>
  <aug>
    <au><snm>Zuech</snm><fnm>R</fnm></au>
    <au><snm>Khoshgoftaar</snm><fnm>TM</fnm></au>
    <au><snm>Wald</snm><fnm>R</fnm></au>
  </aug>
  <source>Journal of Big Data</source>
  <publisher>Springer</publisher>
  <pubdate>2015</pubdate>
</bibl>

<bibl id="B28">
  <title><p>{A Study on Zero-Day Attacks}</p></title>
  <aug>
    <au><snm>Akshaya</snm><fnm>S</fnm></au>
    <au><cnm>others</cnm></au>
  </aug>
  <pubdate>2019</pubdate>
</bibl>

<bibl id="B29">
  <title><p>{Survey of intrusion detection systems: techniques, datasets and
  challenges}</p></title>
  <aug>
    <au><snm>Khraisat</snm><fnm>A</fnm></au>
    <au><snm>Gondal</snm><fnm>I</fnm></au>
    <au><snm>Vamplew</snm><fnm>P</fnm></au>
    <au><snm>Kamruzzaman</snm><fnm>J</fnm></au>
  </aug>
  <source>Cybersecurity</source>
  <publisher>Springer</publisher>
  <pubdate>2019</pubdate>
</bibl>

<bibl id="B30">
  <title><p>{The industrial control system cyber defence triage
  process}</p></title>
  <aug>
    <au><snm>Cook</snm><fnm>A</fnm></au>
    <au><snm>Janicke</snm><fnm>H</fnm></au>
    <au><snm>Smith</snm><fnm>R</fnm></au>
    <au><snm>Maglaras</snm><fnm>L</fnm></au>
  </aug>
  <source>Computers \& Security</source>
  <publisher>Elsevier</publisher>
  <pubdate>2017</pubdate>
</bibl>

<bibl id="B31">
  <title><p>{Process-aware model based IDSs for industrial control systems
  cybersecurity: approaches, limits and further research}</p></title>
  <aug>
    <au><snm>Escudero</snm><fnm>C</fnm></au>
    <au><snm>Sicard</snm><fnm>F</fnm></au>
    <au><snm>Zama{\"\i}</snm><fnm>{\'E}</fnm></au>
  </aug>
  <source>ETFA</source>
  <pubdate>2018</pubdate>
</bibl>

<bibl id="B32">
  <title><p>{Distributed intrusion detection system using semantic-based rules
  for scada in smart grid}</p></title>
  <aug>
    <au><snm>Mohan</snm><fnm>SN</fnm></au>
    <au><snm>Ravikumar</snm><fnm>G</fnm></au>
    <au><snm>Govindarasu</snm><fnm>M</fnm></au>
  </aug>
  <source>T\&D</source>
  <pubdate>2020</pubdate>
</bibl>

<bibl id="B33">
  <title><p>{Efficient modelling of ics communication for anomaly detection
  using probabilistic automata}</p></title>
  <aug>
    <au><snm>Matou{\v{s}}ek</snm><fnm>P</fnm></au>
    <au><snm>Havlena</snm><fnm>V</fnm></au>
    <au><snm>Hol{\'\i}k</snm><fnm>L</fnm></au>
  </aug>
  <source>IM</source>
  <pubdate>2021</pubdate>
</bibl>

<bibl id="B34">
  <title><p>{Fuzzy automaton as a detection mechanism for the multi-step
  attack}</p></title>
  <aug>
    <au><snm>Almseidin</snm><fnm>M</fnm></au>
    <au><snm>Piller</snm><fnm>I</fnm></au>
    <au><snm>Al Kasassbeh</snm><fnm>M</fnm></au>
    <au><snm>Kovacs</snm><fnm>S</fnm></au>
  </aug>
  <source>IJASEIT</source>
  <pubdate>2019</pubdate>
</bibl>

<bibl id="B35">
  <title><p>{SCAPY-A powerful interactive packet manipulation
  program}</p></title>
  <aug>
    <au><snm>Rohith</snm><fnm>R</fnm></au>
    <au><snm>Moharir</snm><fnm>M</fnm></au>
    <au><snm>Shobha</snm><fnm>G</fnm></au>
    <au><cnm>others</cnm></au>
  </aug>
  <source>ICNEWS</source>
  <pubdate>2018</pubdate>
</bibl>

<bibl id="B36">
  <title><p>{CICFlowMeter}</p></title>
  <aug>
    <au><snm>Lashkari</snm><fnm>AH</fnm></au>
    <au><snm>Zang</snm><fnm>Y</fnm></au>
    <au><snm>Owhuo</snm><fnm>G</fnm></au>
    <au><snm>Mamun</snm><fnm>MSI</fnm></au>
    <au><snm>Gil</snm><fnm>GD</fnm></au>
  </aug>
  <publisher>Github</publisher>
  <pubdate>2017</pubdate>
</bibl>

<bibl id="B37">
  <title><p>{An anomaly detection mechanism for IEC 60870-5-104}</p></title>
  <aug>
    <au><snm>Grammatikis</snm><fnm>PR</fnm></au>
    <au><snm>Sarigiannidis</snm><fnm>P</fnm></au>
    <au><snm>Sarigiannidis</snm><fnm>A</fnm></au>
    <au><snm>Margounakis</snm><fnm>D</fnm></au>
    <au><snm>Tsiakalos</snm><fnm>A</fnm></au>
    <au><snm>Efstathopoulos</snm><fnm>G</fnm></au>
  </aug>
  <source>MOCAST</source>
  <pubdate>2020</pubdate>
</bibl>

<bibl id="B38">
  <title><p>{Anomaly Detection of ICS Communication Using Statistical
  Models}</p></title>
  <aug>
    <au><snm>Burgetov{\'a}</snm><fnm>I</fnm></au>
    <au><snm>Matou{\v{s}}ek</snm><fnm>P</fnm></au>
    <au><snm>Ry{\v{s}}av{\`y}</snm><fnm>O</fnm></au>
  </aug>
  <source>CNSM</source>
  <pubdate>2021</pubdate>
</bibl>

<bibl id="B39">
  <title><p>{A Comparison of Unsupervised Learning Algorithms for Intrusion
  Detection in IEC 104 SCADA Protocol}</p></title>
  <aug>
    <au><snm>Anwar</snm><fnm>M</fnm></au>
    <au><snm>Borg</snm><fnm>A</fnm></au>
    <au><snm>Lundberg</snm><fnm>L</fnm></au>
  </aug>
  <source>ICMLC</source>
  <pubdate>2021</pubdate>
</bibl>

<bibl id="B40">
  <title><p>{Status of the National Implementation of the NC RfG in
  Germany}</p></title>
  <aug>
    <au><snm>Scheben</snm><fnm>F</fnm></au>
    <au><snm>Genzmer</snm><fnm>K</fnm></au>
    <au><snm>Mohrdieck</snm><fnm>JM</fnm></au>
    <au><snm>M{\"o}ller</snm><fnm>J</fnm></au>
  </aug>
  <source>NEIS Conference 2016</source>
  <pubdate>2017</pubdate>
</bibl>

<bibl id="B41">
  <title><p>Snort 2.1 intrusion detection</p></title>
  <aug>
    <au><snm>Caswell</snm><fnm>B</fnm></au>
    <au><snm>Beale</snm><fnm>J</fnm></au>
  </aug>
  <publisher>Alibris United States: Elsevier</publisher>
  <pubdate>2004</pubdate>
</bibl>

<bibl id="B42">
  <title><p>{Improving the performance of the intrusion detection systems by
  the machine learning explainability}</p></title>
  <aug>
    <au><snm>Dang</snm><fnm>QV</fnm></au>
  </aug>
  <source>International Journal of Web Information Systems</source>
  <publisher>Emerald Publishing Limited</publisher>
  <pubdate>2021</pubdate>
</bibl>

<bibl id="B43">
  <title><p>{Measuring the quality of explanations: the system causability
  scale (SCS)}</p></title>
  <aug>
    <au><snm>Holzinger</snm><fnm>A</fnm></au>
    <au><snm>Carrington</snm><fnm>A</fnm></au>
    <au><snm>M{\"u}ller</snm><fnm>H</fnm></au>
  </aug>
  <source>KI-K{\"u}nstliche Intelligenz</source>
  <publisher>Springer</publisher>
  <pubdate>2020</pubdate>
</bibl>

<bibl id="B44">
  <title><p>{Automata, dynamical systems, and groups}</p></title>
  <aug>
    <au><snm>Grigorchuk</snm><fnm>RI</fnm></au>
    <au><snm>Nekrashevych</snm><fnm>VV</fnm></au>
    <au><snm>Sushchansky</snm><fnm>VI</fnm></au>
  </aug>
  <source>Trudy Matematicheskogo Instituta Imeni VA Steklova</source>
  <publisher>Russian Academy of Sciences</publisher>
  <pubdate>2000</pubdate>
</bibl>

<bibl id="B45">
  <title><p>{Towards model-based anomaly detection in network communication
  protocols}</p></title>
  <aug>
    <au><snm>Bieniasz</snm><fnm>J</fnm></au>
    <au><snm>Sapiecha</snm><fnm>P</fnm></au>
    <au><snm>Smolarczyk</snm><fnm>M</fnm></au>
    <au><snm>Szczypiorski</snm><fnm>K</fnm></au>
  </aug>
  <source>ICFSP</source>
  <pubdate>2016</pubdate>
</bibl>

<bibl id="B46">
  <title><p>{libalf: The automata learning framework}</p></title>
  <aug>
    <au><snm>Bollig</snm><fnm>B</fnm></au>
    <au><snm>Katoen</snm><fnm>JP</fnm></au>
    <au><snm>Kern</snm><fnm>C</fnm></au>
    <au><snm>Leucker</snm><fnm>M</fnm></au>
    <au><snm>Neider</snm><fnm>D</fnm></au>
    <au><snm>Piegdon</snm><fnm>DR</fnm></au>
  </aug>
  <source>CAV</source>
  <pubdate>2010</pubdate>
</bibl>

<bibl id="B47">
  <title><p>{Towards an Approach to Contextual Detection of Multi-Stage Cyber
  Attacks in Smart Grids}</p></title>
  <aug>
    <au><snm>Sen</snm><fnm>{\"O}</fnm></au>
    <au><snm>Velde</snm><fnm>D</fnm></au>
    <au><snm>Wehrmeister</snm><fnm>KA</fnm></au>
    <au><snm>Hacker</snm><fnm>I</fnm></au>
    <au><snm>Henze</snm><fnm>M</fnm></au>
    <au><snm>Andres</snm><fnm>M</fnm></au>
  </aug>
  <source>SEST</source>
  <pubdate>2021</pubdate>
</bibl>

<bibl id="B48">
  <title><p>{On Using Contextual Correlation to Detect Multi-Stage Cyber
  Attacks in Smart Grids}</p></title>
  <aug>
    <au><snm>Sen</snm><fnm>{\"O}</fnm></au>
    <au><snm>Velde</snm><fnm>D</fnm></au>
    <au><snm>Wehrmeister</snm><fnm>K</fnm></au>
    <au><snm>Hacker</snm><fnm>I</fnm></au>
    <au><snm>Henze</snm><fnm>M</fnm></au>
    <au><snm>Andres</snm><fnm>M</fnm></au>
  </aug>
  <source>SEGAN</source>
  <pubdate>2022</pubdate>
</bibl>

<bibl id="B49">
  <title><p>IEC104 Client Utility - Metasploit</p></title>
  <aug>
    <au><cnm>infosecmatter</cnm></au>
  </aug>
  <url>https://www.infosecmatter.com/metasploit-module-library/?mm=auxiliary/client/iec104/iec104</url>
</bibl>

<bibl id="B50">
  <title><p>{Investigating Man-in-the-Middle-based False Data Injection in a
  Smart Grid Laboratory Environment}</p></title>
  <aug>
    <au><snm>Sen</snm><fnm>{\"O}</fnm></au>
    <au><snm>Van Der Veldc</snm><fnm>D</fnm></au>
    <au><snm>Linnartz</snm><fnm>P</fnm></au>
    <au><snm>Hacker</snm><fnm>I</fnm></au>
    <au><snm>Henze</snm><fnm>M</fnm></au>
    <au><snm>Andres</snm><fnm>M</fnm></au>
    <au><snm>Ulbig</snm><fnm>A</fnm></au>
  </aug>
  <source>ISGT Europe</source>
  <pubdate>2021</pubdate>
</bibl>

<bibl id="B51">
  <title><p>ProfiShark 1G+ Datasheet</p></title>
  <aug>
    <au><snm>B.V.</snm><fnm>PH</fnm></au>
  </aug>
  <url>https://www.profitap.com/wp-content/uploads/ProfiShark-1G-Plus-Datasheet.pdf</url>
</bibl>

<bibl id="B52">
  <title><p>{Classification assessment methods}</p></title>
  <aug>
    <au><snm>Tharwat</snm><fnm>A</fnm></au>
  </aug>
  <source>Applied Computing and Informatics</source>
  <publisher>Emerald Publishing Limited</publisher>
  <pubdate>2020</pubdate>
</bibl>

<bibl id="B53">
  <title><p>{OSCIDS: An Ontology based SCADA Intrusion Detection
  Framework.}</p></title>
  <aug>
    <au><snm>Al Balushi</snm><fnm>A</fnm></au>
    <au><snm>McLaughlin</snm><fnm>K</fnm></au>
    <au><snm>Sezer</snm><fnm>S</fnm></au>
  </aug>
  <source>SECRYPT</source>
  <pubdate>2016</pubdate>
</bibl>

<bibl id="B54">
  <title><p>{A cybersecurity detection framework for supervisory control and
  data acquisition systems}</p></title>
  <aug>
    <au><snm>Cruz</snm><fnm>T</fnm></au>
    <au><snm>Rosa</snm><fnm>L</fnm></au>
    <au><snm>Proen{\c{c}}a</snm><fnm>J</fnm></au>
    <au><snm>Maglaras</snm><fnm>L</fnm></au>
    <au><snm>Aubigny</snm><fnm>M</fnm></au>
    <au><snm>Lev</snm><fnm>L</fnm></au>
    <au><snm>Jiang</snm><fnm>J</fnm></au>
    <au><snm>Sim{\~o}es</snm><fnm>P</fnm></au>
  </aug>
  <source>IEEE Transactions on Industrial Informatics</source>
  <publisher>IEEE</publisher>
  <pubdate>2016</pubdate>
</bibl>

<bibl id="B55">
  <title><p>{Exploiting bro for intrusion detection in a SCADA
  system}</p></title>
  <aug>
    <au><snm>Udd</snm><fnm>R</fnm></au>
    <au><snm>Asplund</snm><fnm>M</fnm></au>
    <au><snm>Nadjm Tehrani</snm><fnm>S</fnm></au>
    <au><snm>Kazemtabrizi</snm><fnm>M</fnm></au>
    <au><snm>Ekstedt</snm><fnm>M</fnm></au>
  </aug>
  <source>CPS-SPC</source>
  <pubdate>2016</pubdate>
</bibl>

<bibl id="B56">
  <title><p>{Multidimensional intrusion detection system for IEC 61850-based
  SCADA networks}</p></title>
  <aug>
    <au><snm>Yang</snm><fnm>Y</fnm></au>
    <au><snm>Xu</snm><fnm>HQ</fnm></au>
    <au><snm>Gao</snm><fnm>L</fnm></au>
    <au><snm>Yuan</snm><fnm>YB</fnm></au>
    <au><snm>McLaughlin</snm><fnm>K</fnm></au>
    <au><snm>Sezer</snm><fnm>S</fnm></au>
  </aug>
  <source>IEEE Transactions on Power Delivery</source>
  <publisher>IEEE</publisher>
  <pubdate>2016</pubdate>
</bibl>

<bibl id="B57">
  <title><p>{Distributed attack detection in a water treatment plant: Method
  and case study}</p></title>
  <aug>
    <au><snm>Adepu</snm><fnm>S</fnm></au>
    <au><snm>Mathur</snm><fnm>A</fnm></au>
  </aug>
  <source>IEEE Transactions on Dependable and Secure Computing</source>
  <publisher>IEEE</publisher>
  <pubdate>2018</pubdate>
</bibl>

<bibl id="B58">
  <title><p>{Runtime semantic security analysis to detect and mitigate
  control-related attacks in power grids}</p></title>
  <aug>
    <au><snm>Lin</snm><fnm>H</fnm></au>
    <au><snm>Slagell</snm><fnm>A</fnm></au>
    <au><snm>Kalbarczyk</snm><fnm>ZT</fnm></au>
    <au><snm>Sauer</snm><fnm>PW</fnm></au>
    <au><snm>Iyer</snm><fnm>RK</fnm></au>
  </aug>
  <source>IEEE Transactions on Smart Grid</source>
  <publisher>IEEE</publisher>
  <pubdate>2016</pubdate>
</bibl>

<bibl id="B59">
  <title><p>{Intrusion detection model of SCADA using graphical
  features}</p></title>
  <aug>
    <au><snm>Wang</snm><fnm>D</fnm></au>
    <au><snm>Feng</snm><fnm>D</fnm></au>
  </aug>
  <source>IAEAC</source>
  <pubdate>2018</pubdate>
</bibl>

</refgrp>
} 

\end{backmatter}
\end{document}